\documentclass[12pt,preprint]{aastex}
\usepackage{psfig}
\newcommand{\beq}{\begin{equation}}
\newcommand{\eeq}{\end{equation}}
\newcommand{\beqn}{\begin{eqnarray}}
\newcommand{\eeqn}{\end{eqnarray}}

\begin{document}

\title{\bf{Data Analysis on the Extra-solar Planets Using Robust
Clustering } }
\author{Ing-Guey Jiang$^a$, Li-Chin Yeh$^b$, Wen-Liang Hung$^b$,
Miin-Shen Yang$^c$}

\affil{{\small $^a$Department of Physics, National Tsing-Hua
 University, Hsin-Chu, Taiwan}\\
{\small $^b$Department of Applied Mathematics, } \\
{\small National Hsinchu University of Education, Hsin-Chu, Taiwan} \\
{\small $^c$Department of Applied Mathematics, } \\
{\small Chung Yuan Christian University, Chung-Li, Taiwan}}


\begin{abstract}

We use both the conventional and more recently developed methods of
cluster analysis to study the data of extra-solar planets.
Using the data set with planetary mass $M$, orbital period $P$,
and orbital eccentricity $e$, we investigate the possible
 clustering in the ${\rm ln} M$, ${\rm ln} P$,
 ${\rm ln} P-{\rm ln} M$, $e$, and ${\rm ln} P-e$ spaces.
There are two main
implications: (1) mass distribution is continuous and (2) orbital
population could be classified into three clusters, which correspond to
the exoplanets in the regimes of tidal, on-going tidal and disc
interaction, respectively.
\end{abstract}

\noindent
 {\bf Key words:} data analysis, cluster analysis,
planetary systems, stellar dynamics.

\section{Introduction}

Astronomers' observational efforts have led to the detection of
more than 160 extra-solar planets (exoplanets). These discoveries
open a new window to astronomy, which could eventually lead to
answers to the fundamental questions about the formation of
planetary systems, including our Solar System. In fact, some of
the work has already been done to address the important questions
about planetary evolution. For example,
Zakamska \& Tremaine (2004) tried to explain the high eccentricities
of exoplanets by possible stellar encounters.
Ji et al. (2002, 2003) have studied the orbital evolutions of
a few known resonant planetary systems.
Veras \& Armitage (2004) proposed the possible mechanisms
to make exoplanets migrate outward.
 Jiang \& Ip (2001)
investigated the origin of orbital elements of the planetary
system of upsilon Andromedae. Yeh \& Jiang (2001) studied the
orbital migration of scattered planets. Jiang \& Yeh (2004a,
2004b) did some analysis on the orbital evolution of systems with
planet-disc interaction.

While the number of detected planets keep increasing, it is
important to study the distributions of their masses, periods, and
other orbital elements. The details of these distributions could
have crucial implications for the formation and evolution of
planetary systems. It is quite common that we use a simple
function, say, a power law, to model the distribution of a
quantity in our system (Tabachnik \& Tremaine 2002).
It also happens that, after a further
study, we find that a single power-law is not good enough and we
make the change to a double-power-law. This process will
eventually help us to understand the properties of the system.

On the other hand, cluster analysis, which is a data analysis tool
to find clusters of a data set with the most similarity in the
same cluster and the most dissimilarity between different
clusters, might also help us to distinguish a double-power-law
from a single power-law for a particular study from another point
of view. When we find that there is only one group (or a huge
number of groups) through cluster analysis, it is likely that a
simple function (say, a single power-law) could be a good
approximation. When there are two or more groups  in  cluster
analysis, we might need a combination of several functions (say,
double-power-law) to model the data set. In this way, we can be
more confident of the results derived from the usual conventional
 process.  Another advantage of cluster analysis is that it
can also be used to study the distributions of multi-variables
easily.

Clustering is a powerful exploratory approach to find groups in
data and to reveal the structure information of a given data
set. It is a data driven procedure to classify a datum in one of a
few classes by looking at proximity and homogeneity in feature
space. Conventional approaches to clustering can be grouped into
two categories namely partitional and hierarchical. The $k$-means
(see MacQueen 1967, Hartigan 1975) algorithm is a popular
example of partitional clustering whereby the data is partitioned
into $k$ classes with the value of $k$ known a priori. Single
linkage algorithm (Gower and Ross 1969, Hartigan 1967) is an
example of hierarchical agglomerative clustering where for $n$
data a hierarchy of clusters ranging from $n$ to $1$ is formed.


The single linkage algorithm was used in asteroid studies
(Zappala et al. 1995) and  in many meteoroid stream searches
(Baggaley \& Galligan 1997, Galligan 2003a, Galligan 2003b).
Zappala et al. (1995) identified the dynamical families from a
sample of 12,487 asteroids. They divided the main asteroid belt
into three different zones (the inner, intermediate, and outer
zone) and study the hierarchical structures of the orbital
families.
Baggaley \& Galligan (1997) first used the
single linkage algorithm to work on meteoroid stream data from the
Advanced Meteoroid Orbit Radar (AMOR). They probed the structure
of the orbital distributions and determined the extent of
dynamical families or clusters in the population. Moreover, in
order to determine a reasonable cut-off level, they also
introduced the combination of the single-linkage method with a
randomization technique.  Galligan (2003a, 2003b) further studied
more recent AMOR data by these methods.

From the previous work, we can see that the cut-off level in the
conventional single linkage method would influence the results of
clustering and therefore the number of clusters. In Section 2, we
first briefly review the single linkage algorithm. We then review
the much more advanced method called the similarity-based
clustering method (SCM) proposed by Yang \& Wu (2004) in Section
3. Some numerical examples and comparisons between the single linkage algorithm
and the SCM are made in Section 4. Section 5 starts our
applications of the single linkage algorithm and the SCM on the
data of exoplanets. The implications of these results on the
formation and evolution of planetary systems will also be given in
Section 5. Section 6 concludes the paper.

\section{The Single Linkage Algorithm}

The clustering process begins with measures of the distance of the
objects from one another. Several measures are available, but we
shall use simple Euclidean distance in most cases. Usually, the
distances can be summarized in a symmetric $n\times n$ matrix
$D=(d_{ij})$, where $d_{ij}$ is the Euclidean distance between
objects $i$ and $j$. 
We illustrate the single linkage algorithm with the
artificial distance matrix given by Johnson \& Wichern (1988)
\[D=\left[\begin{array}{ccccc}
 0  &    &   &   & \\
 9  & 0  &   &   & \\
 3  & 7  & 0 &   & \\
 6  & 5  & 9 & 0 & \\
 11 & 10 & 2 & 8 & 0
\end{array}\right].
\] 
The minimum distance occurs between objects $3$ and $5$, so the
first cluster, labelled as (35), 
will consist of those objects. The second matrix of
distances is formed by deleting the rows and columns of $D$
corresponding to objects $3$ and $5$, and adding a row and column
with the distances of the remaining objects from the cluster (35).
Those distances are found from the rule
$$d_{(35),1}=\min\{d_{31}, d_{51}\}=\min\{3,11\}=3$$
$$d_{(35),2}=\min\{d_{32}, d_{52}\}=\min\{7,10\}=7$$
$$d_{(35),4}=\min\{d_{34}, d_{54}\}=\min\{9,8\}=8.$$
The minimum distance in the second matrix is $d_{(35),1}=3$ and we
merge cluster (1) with cluster (35) to get the next cluster,
(135). Next, we form a third distance matrix: for it we note that
$$d_{(135),2}=\min\{d_{(35),2}, d_{12}\} =\min\{7,9\}=7$$ 
$$d_{(135),4}=\min\{d_{(35),4}, d_{14}\} =\min\{8,6\}=6.$$
The minimum distance in the third matrix is $d_{42}=5$ and we
merge objects 4 and 2 to get the cluster (24). At this time we
have two distinct clusters, (135) and (24). Their minimum distance
is
$$d_{(135),(24)}=\min\{d_{(135),2}, d_{(135),4}\}=\min\{7,6\}=6.$$
Therefore, clusters (135) and (24) are merged to form a single
cluster of all five objects, (12345), when the minimum distance
reaches 6.

The clusters are illustrated in the dendrogram of Fig. 1. The
dendrogram is formed by plotting the minimum distances for each
cluster in a tree configuration leading to the single cluster of
all $5$ objects. The objects should be ordered so that the
branches of the dendrogram stand alone without crossing. The
dendrogram clearly suggests that the sample may contain two sets
of objects: $(135)$ and $(24)$.

In general, single linkage algorithm,
which combines the
original $n$ single-object clusters hierarchically into one
cluster of $n$ objects, cab be summarized as:

\vskip 0.1truein \noindent
 {\bf Single Linkage Algorithm}

\begin{itemize}
\item[S1.] Start with $n$ single-object clusters.
The distances are described by the initial distance matrix
$D=(d_{ij})$.

\item[S2.] Search for the nearest pair of
clusters from the distance matrix. When 
there are more than one candidate pairs,
randomly pick one of them.
Let the distance between this pair of clusters, $U$
and $V$, be $d_{UV}$, where
$$d_{UV}=\min_{u \in U, v\in V}\{d_{uv}\}.$$

 \item[S3.] Merge clusters $U$ and $V$. Label the newly formed
 cluster as $(UV)$. Form a new distance matrix by (i)
 deleting the rows and columns corresponding to clusters $U$ and
 $V$ and (ii) adding a row and column giving the distances between
 cluster $(UV)$ and the remaining clusters.
The distances between
cluster $(UV)$ and any other cluster $W$ are computed by
 $$d_{(UV)W}=\min\{d_{UW},d_{VW}\},$$
 where $d_{UW}$ and $d_{VW}$ are the distances
 between clusters $U$ and $W$ and clusters $V$ and $W$, as defined in S2.

\item[S4.] Repeat S2 and S3. All objects
will be in a single cluster at the termination of the algorithm.
Record the identity of clusters that are merged and the levels at
which the mergers take place.
  \end{itemize}

The above clustering results of single linkage algorithm can be
graphically displayed in the form of a dendrogram, which
shows the clustering structure at various levels
of the hierarchy. The branches in the tree represent clusters. The
branches come together (merge) at nodes whose positions along a
distance axis indicate the level at which the fusions occur.
Therefore, the vertical axis
indicates the distance and
the horizontal axis 
simply shows the data identity numbers.

Please note that the dendrogram plotted in Baggaley \& Galligan
(1997) is slightly different from what we just described here. In
their plot, the meaning of the horizontal axis is different from
ours. Due to the huge number of orbits, their horizontal axis
shows the scale of number of clustering members. The meaning of
their vertical axis is the same as ours.

\section{The Similarity-Based Clustering Method}

Let the data set be $X=\{x_1,\cdots,x_n\}$ where $x_j$ is a
feature vector in the $s$-dimensional Euclidean space $\Re^s$ and
$c$ is the specified number of clusters. Yang and Wu (2004)
considered maximizing the objective function $J(z)$ with
 \begin{eqnarray}
 J(z)=\sum_{i=1}^c\sum_{j=1}^n \Big(\exp -\frac{||x_j-z_i||^2}{\beta}\Big)^\gamma,
 \end{eqnarray}
where $\exp(-\frac{||x_j-z_i||^2}{\beta})$ is the similarity
measure between $x_j$ and the $i$th cluster center $z_i$,
$||x_j-z_i||$ is the Euclidean norm, $\gamma>0$ and
$$\beta=\frac{\sum_{j=1}^n ||x_j-\bar x||^2}{n},\,\,\mbox{where}\,\,
\bar x=\frac{1}{n}\sum_{j=1}^n x_j.$$

 Since the clustering result is influenced by $\gamma$,  Yang and Wu (2004)
 proposed correlation comparison algorithm (CCA)
 to select $\gamma$. According to the fact that the parameter
 $\gamma$ controls the location of
 the peaks of $J(z)$, they considered the total similarity function
  $\tilde J(x_k)_{\gamma_m}$
 for each data point $x_k$ with
\begin{eqnarray*}
 \tilde J(x_k)_{\gamma_m}=\sum_{j=1}^n \Big(\exp-\frac{||x_j-x_k||^2}{\beta}\Big)^{\gamma_m},~
 k=1,\cdots,n,
 \end{eqnarray*}
where $\gamma_m=5m$, $m=1,2,3,\cdots$. The correlation between the
values of $\tilde J(x_k)_{\gamma_m}$ and
  $\tilde J(x_k)_{\gamma_{m+1}}$
 are calculated. That is, CCA is based on a correlation comparison procedure with
 ``$\gamma_1=5, \gamma_2=10$'', ``$\gamma_2=10,
 \gamma_3=15$'', ``$\gamma_3=15, \gamma_4=20$'', $\cdots$ etc. The CCA then is summarized as
follows:

\vskip 0.1truein
\noindent
 {\bf Correlation Comparison Algorithm (CCA)}

\begin{itemize}
\item[S1.] Set $m=1$ and give a threshold $\epsilon_1$.

\item[S2.] Calculate the correlation of the values of
 $\tilde J(x_k)_{\gamma_m}$ and $\tilde J(x_k)_{\gamma_{m+1}}$.

 \item[S3.] If the correlation is greater than or equal to the threshold $\epsilon_1$

\item[] THEN choose $\gamma_m$ to be the estimate of $\gamma$;

 \item[] ELSE $m=m+1$ and GOTO S2.
 \end{itemize}

Since Yang and Wu (2004)
suggested a threshold around $0.97\sim 0.999$,
we choose 0.99 for the threshold in this paper. After the parameter
$\gamma$ is estimated using CCA, the next step is to find a $z_i$
that maximizes the SCM objective function $J(z)$. Differentiating
$J(z)$ with respect to all $z_i$, we obtain
\begin{eqnarray}
\frac{d J_(z)}{dz_i}=\sum_{j=1}^n
2\frac{\gamma}{\beta}(x_j-z_i)\Big(\exp-\frac{||x_j-z_i||^2}{\beta}\Big)^\gamma
\end{eqnarray}
and set (2) to zero. The necessary condition that maximizes
$J_(z)$ is
\begin{eqnarray}
 z_i=\frac{\sum_{j=1}^n x_j\Big(\exp(-\frac{||x_j-z_i||^2}{\beta})\Big)^\gamma}
 {\sum_{j=1}^n\Big(\exp(-\frac{||x_j-z_i||^2}{\beta})\Big)^\gamma}.
\end{eqnarray}
This necessary condition can be decomposed into two conditions.
First, we take the similarity relation $S(x_j,z_i)$ with
\begin{eqnarray}
S_{ij}=S(x_j,z_i)=\exp\Big(-\frac{||x_j-z_i||^2}{\beta}\Big)
\end{eqnarray}
and then the necessary condition (3) becomes
\begin{eqnarray}
 z_i=\frac{\sum_{j=1}^nS_{ij}^\gamma x_j}{\sum_{j=1}^n S_{ij}^\gamma}.
\end{eqnarray}
 This forms the similarity clustering algorithm (SCA). Thus, after the
CCA is implemented to get an estimate $\gamma$, the SCA will be
used to find the peaks of the SCM objective function is then
summarized as follows:

 \vskip 0.1truein
\noindent
 {\bf Similarity Clustering Algorithm (SCA)}

\begin{itemize}
\item[] Initialize $z_i^{(0)},i=1,\cdots,c$ and give $\epsilon$;

\item[] Set iteration counter $\ell=0$;

\item[S1.] Estimate $S_{ij}^{(\ell+1)}$ by Eq.(4);

\item[S2.] Estimate $z_i^{(\ell+1)}$ by Eq.(5);

\item[] Increment $\ell$; Until $\displaystyle\max_i
||z_i^{(\ell+1)}-z_i^{(\ell)}||<\epsilon$.
 \end{itemize}

When one processes SCA, all the cluster centers, $z_i$, will change
positions for each iteration.
If the data set has only one peak on the SCM objective
function, all the centers will gradually centralize to that unique peak.
In this case, we will claim there is only one cluster for this data set.
When the data set has more than one peak on the SCM objective
function,
we can randomly give more initial cluster
centers to process SCA and these centers will then
centralize to the peaks of the SCM objective function. The
problem here is what kind of the initialization can guarantee that
all peaks (clusters) will be found simultaneously. To solve
this problem, Yang and Wu (2004) suggested to set all data points
to be the initial centers (i.e.,
$z^{(0)}=(z_1^{(0)},\cdots,z_n^{(0)})=(x_1,\cdots,x_n))$. They
successfully showed that all peaks (clusters) will be found for
this initialization.

Next, we present an example to illustrate the CCA and SCA. In the
data set of Fig. 2, there are one large cluster and two small
clusters.
By the CCA, we find that $\gamma=10$ is a good estimate. Then we
process SCA with this data set by initializing
$z^{(0)}=(z_1^{(0)},\cdots,z_n^{(0)})=(x_1,\cdots,x_n)$.
We show the positions of these
cluster centers after 1, 5 and 10 iterations as the full circles
in Fig. 3 and the left of Fig. 4, respectively.
The final convergent positions of all
cluster centers are shown as the full circles in the right of Fig. 4.
Please note that all the data points, $(x_1,\cdots,x_n)$, are also plotted
as the dots in Fig. 3 and 4.
It is obvious that there are only three full circles because all cluster
centers centralize to the three peaks of the SCM objective function.
There are three clusters for this data set by the view of sight.
However, a precise method to
determine the cluster number
from these final $n$ cluster centers should be
provided. We use the single linkage algorithm with the final
positions of all cluster centers to find the optimal cluster number $c^*$.
The result is shown as the dendrogram
in the left panel of Fig. 5. This dendrogram clearly indicates
that there are three well separated clusters and hence
the optimal cluster number $c^*=3$.
At the same time the
identified clusters will be found. The identified clusters are shown
in the right panel of Fig. 5.

Therefore, we can completely cluster the data set in Fig. 2 into
three clusters shown in the right panel of Fig. 5.
The process includes that:
(i) process CCA to estimate the parameter $\gamma$,
(ii) process SCA with
$z^{(0)}=(z_1^{(0)},\cdots,z_n^{(0)})=(x_1,\cdots,x_n)$, and
(iii) process the single linkage algorithm with the final
positions of $n$ cluster centers
to find the optimal cluster number $c^*$ and identify these $c^*$
clusters. This forms the structure of SCM which is summarized as
follows:

 \vskip 0.1truein
\noindent
 {\bf Similarity-Based Clustering Method (SCM)}

\begin{itemize}
 \item[S1.] Estimate $\gamma$ using CCA;
 \item[S2.] Process SCA with
$z^{(0)}=(z_1^{(0)},\cdots,z_n^{(0)})=(x_1,\cdots,x_n)$;
 \item[S3.] Process the single linkage algorithm
with the final $n$ cluster centers;
 \item[S4.] Find the optimal cluster number $c^*$ according to the
 dendrogram;
  \item[S5.] Identify these $c^*$ clusters.
 \end{itemize}

\section{The Numerical Examples and Comparisons}

In this section, we consider the bivariate normal mixtures of
three classes in order to assess the performance of the single
linkage algorithm and SCM. Let $N_2(\mathbf{a},\mathbf{\Sigma})$
represent the bivariate normal with mean vector $\mathbf{a}$ and
covariance matrix $\mathbf{\Sigma}$, we consider the
random sample of data drawn from
 $$\alpha_1 N_2({\mathbf{a}}_1,{\mathbf{\Sigma}_1})+
 \alpha_2 N_2({\mathbf{a}}_2,{\mathbf{\Sigma}_2})+
 \alpha_3 N_2({\mathbf{a}}_3,{\mathbf{\Sigma}_3})$$
 with $\alpha_1+\alpha_2+\alpha_3=1$, $0<\alpha_i<1,~i=1,2,3$.
 We design various bivariate normal
mixture distributions shown in Table 1.  In each test, we consider
the sample size $n=200$.

\vskip 0.1truein
 {\scriptsize
 \centerline{{\bf Table 1.} Various Bivariate Normal Mixture Distributions
for the Three Tests}
  \begin{center}
       \begin{tabular}{|c|c|c|c|c|c|c|c|c|c|}\hline
Test &  $\alpha_1$ & ${\mathbf{a}}_1$ & ${\mathbf{\Sigma}_1}$ &$\alpha_2$ & 
${\mathbf{a}}_2$
& ${\mathbf{\Sigma}_2}$  &$\alpha_3$ & ${\mathbf{a}}_3$ & 
${\mathbf{\Sigma}_3}$  \\\hline \hline
 1    & $0.2$ & $(1,1)$ & $ \left(
                            \begin{array}{cc}
                               1.2 & 0 \\
                               0   & 1.2
                               \end{array}\right)$
         &$0.3$ & $(5.7,5.7)$ &
                        $\left(
                           \begin{array}{cc}
                               1 & 0 \\
                               0   & 1
                               \end{array}\right)$
        &$0.5$& $(9,9)$ &
                        $\left(

                           \begin{array}{cc}
                               1.5 & 0 \\
                               0   & 1.5
                               \end{array}\right)$\\ \hline
 2    & $0.2 $& $(1,1)$ & $
                        \left(
                           \begin{array}{cc}
                               2 & 1.5 \\
                             1.5 & 2
                               \end{array}\right)$&
          $0.3$ & $(5,5)$ &
                           $\left(
                           \begin{array}{cc}
                               3 & 0.5 \\
                             0.5 & 3
                               \end{array}\right)$&
           $0.5$ & $(10,10)$ &
                        $\left(
                           \begin{array}{cc}
                               2 & 1 \\
                               1 & 2
                               \end{array}\right)$
                                \\ \hline
 3    & $0.2$ & $(1,3)$ &
                        $\left(
                           \begin{array}{cc}
                               2   & 1.5 \\
                               1.5 & 4
                               \end{array}\right)$
          &$0.3$ &  $(5,7)$  &
                        $\left(
                           \begin{array}{cc}
                               3 & 0.5 \\
                             0.5 & 5
                               \end{array}\right)$
           & $0.5$&  $(9,11)$ & $\left(
                           \begin{array}{cc}
                               6 & 1 \\
                               1 & 4
                               \end{array}\right)$
                                \\\hline
     \end{tabular}
    \end{center}
    }

\normalsize

 \vskip 0.2truein

The data set generated for Test 1 is shown in Fig. 6. From Fig. 6,
it seems there are 2 or 3 clusters. Thus, a precise method to
determine the cluster number is necessary. We process the single
linkage algorithm and SCM with this data set. The clustering
results are shown in Fig. 7.
In the left panel, it is clear that different cut-off level of distances
would give different number of clusters for the single
linkage algorithm's result.
In the right panel of
Fig. 7, SCM clearly indicates that there are three well-separated
clusters. Furthermore, the cluster centers are $(1.2784, 1.3323),
(5.9372, 5.7933), (9.4804, 9.0544)$. These results reflect the
original structure of the data set designed for Test 1.

Fig. 8 shows the data set generated for Test 2. From Fig. 8, the
cluster number seems to be 3. In the same way, we implement the
single linkage algorithm and SCM on this data set.  The dendrogram
of the single linkage algorithm cannot classify this data set well.
But the SCM clearly indicates that there are three well-separated
clusters. Furthermore, the cluster centers are $(1.2096, 1.3824),
(4.9835, 4.6876), (10.6420, 10.3181)$. These results reflect the
structure of the data set designed for Test 2.

Fig. 10 shows the data set generated for Test 3.
It is difficult to know the cluster number from Fig. 10.
We also implement
the single linkage algorithm and SCM on this data set.  The
dendrogram of the single linkage algorithm looks very complicated.
However, the SCM clearly indicates that there are
three well-separated clusters. Furthermore, the cluster centers
are $(1.2373, 3.8693), (5.3338, 7.2524), (9.3596, 11.2041)$. These
results reflect the original structure of
the data set designed for Test 3.

For all the above three tests, SCM gave the optimal cluster number
$c^*=3$ which is exactly the actual cluster number but the single linkage
algorithm fails.
It shows the superiority of SCM over the single
linkage algorithm. As supported by these experiments, SCM produces
satisfactory results with the artificially generated data.

\section{The Results}

In this section,
both the single linkage algorithm and the SCM will be used to cluster
the data of extrasolar planets. Our data is from the Extrasolar
Planets Catalog maintained by Jean Schneider
(http://cfa-www.harvard.edu/planets/catalog.html). We use the data
of April 2005 and exclude the incomplete ones. There are 143
planets (see the Appendix) available for our work and each of them
have the values of $M {\rm sin}~i $, orbital period $P$,
semi-major axis $a$ and also orbital eccentricity $e$, where $M$
is the planetary mass and $i$ is the unknown orbital inclinations.
We set ${\rm sin}~i$ = 1, for simplicity (Trilling et al. 2002).
This assumption shall not have a significant impact on the results
of our clustering analysis because it is not the absolute mass
that matters but the structures of the mass distribution.

\subsection{The Planetary Masses and Periods}

Since there is a possible mass-period correlation for extrasolar
planets (Zucker \& Mazeh 2002, P\"atzold \& Rauer 2002,
Jiang et al. 2003), it should be interesting to see whether there are groups
for extrasolar planets on the ${\rm ln} P$-${\rm ln} M$ plane. The
results of this clustering analysis would give hints about the
formation conditions of these planets.

To investigate the problem step by step, we first study the
possible clustering on the space of ${\rm ln} M$. The dendrogram
illustrating the hierarchical clustering for ${\rm ln} M$ space is
shown in the left panel of Fig. 12. From the dendrogram, we find
that clustering (or mergers) takes place at different levels. For
example, there are two clusters with the distance of about 0.48.
One of them can actually be divided into two with the distance of
about 0.38. Thus, the number of clusters keeps increasing and
becomes very large when a smaller distance is chosen as the
condition to group the data. Consequently, it is difficult to
identity the final configuration of clusters. 
Please note that because there are too many data points, the 
data identity numbers below the horizontal axis of all dendrograms 
in this section are not clear. They can be ignored here. 
To obtain a better
partition for this data, we use the SCM to do the analysis.
The corresponding dendrogram
is shown in the right panel of Fig. 12. From this dendrogram, 
the vertical axis indicates that the
largest level is at $2.4\times 10^{-4},$ which is less than our
chosen $\epsilon$ in SCA (We set $\epsilon=0.001$.). It means
that the minimum distance between pairs of clusters is negligible.
Therefore, all objects are merged to form a single cluster. That
is, by SCM, the data of discovered exoplanets shall be regarded as
a continuous distribution in the $\rm{ln} M$ space.

Indeed, the histogram of the exoplanets in the $\rm{ln} M$ space
in Fig. 13 indicates that there is only one strong peak around
${\ln} M=0$ and the whole distribution seems to be continuous.
Thus, the Jupiter-mass exoplanets dominate the population. Of
course, there are many factors that influence the mass
distribution. For example, the observational selection effect and
perhaps also the orbital instability of close-in massive planets.
Nevertheless, the clustering result showing there is only one
group statistically confirms that there is no particular
constraint on the possible planetary masses except their
boundaries, i.e. the maximum and minimum masses.

We then study the clustering on the ${\rm ln} P$ space. In the
same way, we implement both the single linkage algorithm and SCM
in the ${\rm ln} P$ space. The left and right panels of Fig. 14
show the dendrogram of these two methods, respectively. From the
left panel of Fig. 14, there are no meaningful well-separated
clusters. But there are two well-separated clusters in the right
panel of Fig. 3. Thus, the SCM shows that there are 2 clusters,
and the corresponding clustering results are shown in Table 2. The
cluster with smaller average value of ${\rm ln} P$ is called
Cluster $P_1$ and the other one is called Cluster $P_2$. The
center of Cluster $P_1$ is at 1.4552 and the center of Cluster
$P_2$ is at 6.5455.

\vskip 0.2truein
 {\small \centerline{ {\bf Table 2.} The clustering results
in the ${\rm ln} P$ space.}

    \begin{center}
       \begin{tabular}{|l|llllllllll|}\hline
  Cluster  & \multicolumn{10}{c|}{ data point no.}  \\\hline
 $P_1$     & 1  & 2  & 3  & 4  & 7  & 8  & 9  & 10 & 11 & 12 \\
           & 13 & 14 & 15 & 16 & 17 & 18 & 19 & 20 & 21 & 22\\
           & 25 & 26 & 27 & 28 & 29 & 30 & 33 & 34 & 35 & 37 \\
           & 38 &    &    &    &    &    &    &    &    &  \\\hline
 $P_2$     & 5  & 6  & 23 & 24 & 31 & 32 & 36 & 39 & 40 & 41 \\
           & 42 & 43 & 44 & 45 & 46 & 47 & 48 & 49 & 50 & 51\\
           & 52 & 53 & 54 & 55 & 56 & 57 & 58 & 59 & 60 & 61\\
           & 62 & 63 & 64 & 65 & 66 & 67 & 68 & 69 & 70 & 71\\
           & 72 & 73 & 74 & 75 & 76 & 77 & 78 & 79 & 80 & 81\\
           & 82 & 83 & 84 & 85 & 86 & 87 & 88 & 89 & 90 & 91\\
           & 92 & 93 & 94 & 95 & 96 & 97 & 98 & 99 &100 &101\\
           &102 &103 &104 &105 &106 &107 &108 &109 &110 &111\\
           &112 &113 &114 &115 &116 &117 &118 &119 &120 &121\\
           &122 &123 &124 &125 &126 &127 &128 &129 &130 &131\\
           &132 &133 &134 &135 &136 &137 &138 &139 &140 &141\\
           &142 &143 &    &    &    &    &    &    &    & \\\hline
           \end{tabular}
    \end{center}
    }
\vskip 0.2truein

 \normalsize
 We also make a histogram of our data in the ${\rm ln}
P$ space. As shown in Fig. 15, for this case, it is difficult to
group the data by eye. The two crosses in Fig. 15 indicate the
cluster centers determined by SCM. They are very close to the
peaks of the histogram, so the result of SCM is indeed reasonable
and correct. Thus, we have statistically shown that there are two
clusters for exoplanets in the ${\rm ln} P$ space. The
distribution of orbital periods is unlikely to be a continuous
function. In fact, there is a {\it statistical} gap between two
continuous distributions.

How could that gap form? There could be two possibilities: (1) it
is difficult to form the planets at particular distances from the
central stars, (2) planetary migrations did happen.
The possible planetary migrations as studied in many
theoretical papers are
good candidates to affect the
results of observed period distribution. However, if there is only
one migration mechanism, the resulting distribution is probably
still continuous, unless that
the rate of this
migration mechanism varies as a sharp function of the orbital period.
If there is a period
for which the migration rate is fastest, we might expect much less systems
would end up there.
Another simple interpretation is that there are two important
migration mechanisms to make the exoplanets have two clusters in
the ${\rm ln} P$ space. Indeed, the center of Cluster $P_1$,
1.4552, is within the regime in which the tidal interaction
with the central star is important (Jiang et al. 2003) and the
center of Cluster $P_2$, 6.5455, is within the regime in which the
disc interaction is important (Armitage et al. 2002, Trilling et
al. 2002). The migrations caused by these two mechanisms, tidal
interaction and disc interaction, might produce the two clusters in our
statistical results.

Next, we consider the clustering on the ${\rm ln} P$-${\rm ln} M$
space. In the same way, we implement both the single linkage
algorithm and the SCM on the data in the ${\rm ln} P$-${\rm ln} M$
space. Fig. 16 shows the dendrogram of these two methods. The
right panel of Fig. 16 clearly indicates that there are three
well-separated clusters. Three clusters are called Cluster $PM_1$,
$PM_2$ and $PM_3$ and their centers are at $({\rm ln} P, {\rm ln}
M)=(1.2750, -0.6980)$, $(2.8093,-0.3710)$ and $(6.1522, 0.4857)$,
respectively. Table 3 shows the members of these three clusters.

\vskip 0.2truein
 {\small  \centerline{ {\bf Table 3.} The
clustering results in the ${\rm ln} P$-${\rm ln} M$ space.}

    \begin{center}
       \begin{tabular}{|l|llllllllll|}\hline
  Cluster  & \multicolumn{10}{c|}{ data point no.}  \\\hline
 $PM_1$    & 1  & 2  & 3  & 7  & 8  & 9  & 10 & 11 & 12 & 13\\
           & 14 & 15 & 16 & 17 & 18 & 19 & 20 & 21 & 22 & 25\\
           & 26 & 27 & 28 & 30 & 33 & 34 &    &    &    & \\\hline
 $PM_2$    & 4  & 29 & 35 & 37 & 39 & 40 & 41 & 42 & 43 & 48\\\hline
 $PM_3$    & 5  & 6  & 23 & 24 & 31 & 32 & 36 & 38 & 44 & 45 \\
           & 46 & 47 & 49 & 50 & 51 & 52 & 53 & 54 & 55 & 56 \\
           & 57 & 58 & 59 & 60 & 61 & 62 & 63 & 64 & 65 & 66\\
           & 67 & 68 & 69 & 70 & 71 & 72 & 73 & 74 & 75 & 76\\
           & 77 & 78 & 79 & 80 & 81 & 82 & 83 & 84 & 85 & 86\\
           & 87 & 88 & 89 & 90 & 91 & 92 & 93 & 94 & 95 & 96\\
           & 97 & 98 & 99 &100 &101 &102 &103 &104 &105 &106\\
           &107 &108 &109 &110 &111 &112 &113 &114 &115 &116\\
           &117 &118 &119 &120 &121 &122 &123 &124 &125 &126\\
           &127 &128 &129 &130 &131 &132 &133 &134 &135 &136\\
           &137 &138 &139 &140 &141 &142 &143 &    &    &    \\\hline
                  \end{tabular}
    \end{center}
    }
\vskip 0.2truein

 \normalsize

 Fig. 17 shows the distribution of exoplanet in the
${\rm ln} P$-${\rm ln} M$ space. The triangles are the members of
Cluster $PM_1$, the open squares are the members of Cluster
$PM_2$, and the full circles are the members of Cluster $PM_3$.
There are two are reasons for the crosses going from the
bottom-left to the top-right. The observational selection effect
explains the absence in the bottom-right corner and the tidal
interaction (Jiang et al. 2003) explains the absence in the
top-left corner.

It is not surprising that there are more than one cluster here as
there are already two clusters in the ${\rm ln} P$ space. From
Fig. 15 and 17, we can see that Cluster $PM_1$ is very much
overlapping Cluster $P_1$ and Cluster $PM_3$ is overlapping
Cluster $P_2$. The reason why the Cluster $PM_2$ exists is that
there is a very massive close-in planet, HD 162020 b (data 29),
with mass ($M sin i$) 13.75 $M_J$ and period 8.42 days. This
planet together with a few other massive close-ins stand out as a
new class. If some of the members of Cluster $PM_2$ fall into the
central stars and thus disappear in the plot, this cluster might
be absorbed by the other two clusters. Cluster $PM_2$ might represent
the temporary group with members falling into the stars in the
near future. Thus, we speculate that this cluster experiences
on-going tidal interaction.

\subsection{Orbital Eccentricities}

To understand more about the orbital properties of these
exoplanets, we study the possible clustering in the eccentricity
$e$ space. We shall also study the clustering of exoplanets'
semi-major axes. Because ${\rm ln} a$ is simply a constant times
${\rm ln} P$, the clustering of the data in the ${\rm ln} a$ space
is the same as in the ${\rm ln} P$ space. Thus, we use  ${\rm ln}
P$ to represent ${\rm ln} a$ and study the clustering in the ${\rm
ln P}-e$ space. In the same way, we implement the single linkage
algorithm and SCM on the data in the $e$ space. Fig. 18 shows the
resulting dendrogram by these two methods. The right panel of Fig.
18 clearly indicates that there are four well-separated clusters
but the left panel fails. Therefore, the SCM shows that there are
4 clusters, and the corresponding clustering results are shown in
Table 4. Furthermore, the cluster centers of Cluster $e_1$, $e_2$,
$e_3$, and $e_4$ are at 0.0486, 0.3177, 0.6562, and 0.9207,
respectively.

\newpage
 {\small  \centerline{ {\bf Table 4.} The
clustering result in the $e$ space.}

    \begin{center}
       \begin{tabular}{|l|llllllllll|}\hline
  Cluster  & \multicolumn{10}{c|}{ data point no.}  \\\hline
 $e_1$     & 1  & 2  & 3  & 4  & 6  & 7  & 8  & 9  & 10 & 11 \\
           & 12 & 13 & 14 & 15 & 16 & 17 & 18 & 19 & 20 & 21\\
           & 22 & 25 & 26 & 27 & 28 & 30 & 33 & 37 & 38 & 39 \\
           & 40 & 42 & 44 & 45 & 46 & 48 & 55 & 56 & 57 & 66\\
           & 68 & 71 & 88 & 96 & 99 &102 &111 &113 &118 &119\\
           &120 &135 &136 &139 &    &    &    &    &    & \\\hline
 $e_2$     & 5  & 23 & 24 & 29 & 31 & 34 & 35 & 36 & 41 & 43 \\
           & 47 & 50 & 52 & 53 & 58 & 59 & 61 & 62 & 63 & 64\\
           & 65 & 67 & 70 & 72 & 73 & 74 & 75 & 76 & 77 & 78\\
           & 80 & 81 & 82 & 83 & 84 & 85 & 86 & 87 & 89 & 90\\
           & 91 & 92 & 93 & 94 & 95 & 97 & 98 &100 &101 &103\\
           &106 &107 &108 &109 &110 &114 &116 &117 &121 &122\\
           &123 &125 &126 &127 &128 &129 &130 &131 &132 &133\\
           &137 &140 &141 &142 &143 &    &    &    \\\hline
 $e_3$     & 32 & 49 & 51 & 54 & 69 & 79 &104 &105 &112 &115\\
           &124 &134 &138 &    &    &    &    &    &    &\\\hline
 $e_4$     & 60 &    &    &    &    &    &    &    &    & \\\hline
                  \end{tabular}
    \end{center}
    }
\vskip 0.2truein

\normalsize

 We also plot the histogram of the data in the $e$
space as shown in Fig. 19. From the histogram, it looks as though
there is more than one cluster but it is difficult to determine
the number of clusters by eye. The result of SCM seems to be
reasonable, particularly after we add the crosses to indicate the
centers of these four clusters. Even though there are four
clusters in the $e$ space, there is only one member in the Cluster
$e_4$ and this planet has a very large orbital eccentricity 0.927.

Finally, we consider the single linkage algorithm and the SCM in
the  ${\rm ln} P-e$ space. In the same way, we implement the
single linkage algorithm and SCM on the data. The left and right
panels of Fig. 20 show the dendrogram of these two methods,
respectively. The right panel clearly indicates that there are
three well-separated clusters.  Table 5 shows the SCM's clustering
results. The cluster centers of Cluster $Pe_1$, $Pe_2$, and $Pe_3$
are given by $({\rm ln} P,e)=(1.2860,0.0446)$, $(2.7573, 0.1206)$,
and $(6.3061, 0.3396)$, respectively.

\newpage

 {\small \centerline{ {\bf Table 5.} The
clustering results in the ${\rm ln} P-e$ space.}

    \begin{center}
       \begin{tabular}{|l|llllllllll|}\hline
  Cluster  & \multicolumn{10}{c|}{ data point no.}  \\\hline
 $Pe_1$    & 1  & 2  & 3  & 7  & 8  & 9  & 10 & 11 & 12 & 13\\
           & 14 & 15 & 16 & 17 & 18 & 19 & 20 & 21 & 22 & 25\\
           & 26 & 27 & 28 & 29 & 30 & 33 & 34 &    &    &    \\\hline
 $Pe_2$    & 4  & 35 & 37 & 38 & 39 & 40 & 41 & 42 & 45 & \\\hline
 $Pe_3$    & 5  & 6  & 23 & 24 & 31 & 32 & 36 & 43 & 44 & 46 \\
           & 47 & 48 & 49 & 50 & 51 & 52 & 53 & 54 & 55 & 56 \\
           & 57 & 58 & 59 & 60 & 61 & 62 & 63 & 64 & 65 & 66\\
           & 67 & 68 & 69 & 70 & 71 & 72 & 73 & 74 & 75 & 76\\
           & 77 & 78 & 79 & 80 & 81 & 82 & 83 & 84 & 85 & 86\\
           & 87 & 88 & 89 & 90 & 91 & 92 & 93 & 94 & 95 & 96\\
           & 97 & 98 & 99 &100 &101 &102 &103 &104 &105 &106\\
           &107 &108 &109 &110 &111 &112 &113 &114 &115 &116\\
           &117 &118 &119 &120 &121 &122 &123 &124 &125 &126\\
           &127 &128 &129 &130 &131 &132 &133 &134 &135 &136\\
           &137 &138 &139 &140 &141 &142 &143 &    &    &    \\\hline
                  \end{tabular}
    \end{center}
    }

\vskip 0.2truein
 \normalsize

In the ${\rm ln} P-e$ space as shown in Fig. 21, there are two
data points in Cluster $Pe_2$ with a particularly large $e$ and
thus form a class. They are data 35 and data 41. We find that
these two are also in the Cluster $PM_2$, so they are likely to be
the temporary cluster, in which some of the members will fall into
their central stars in the near future. In general, the exoplanet
data in the ${\rm ln} P-e$ space shows that there is a strong
eccentricity-period correlation. That is, the planets with larger
orbital periods might have larger eccentricities.


\section{Speculations and Implications}

To statistically investigate the properties of the distributions
of the mass, period, and orbits of exoplanets,
we have used both the conventional method, single-linkage algorithm,
and the advanced SCM of Cluster Analysis to
study the possible clusters of the exoplanets in the ${\rm ln} M$,
${\rm ln} P$, ${\rm ln} P-{\rm ln} M$, $e$, and ${\rm ln} P-e$ space.
In general, the SCM gives very good and reasonable results.

We find that there is only one cluster in the ${\rm ln} M$ space,
so a continuous mass function (Tabachnik and Tremaine 2002) would
be a good approximation. We find that there are two clusters in
the ${\rm ln} P$ space and this could be due to two migration
mechanisms; tidal and disc interactions. In addition to the two
clusters associated with the above two mechanisms, there is one
more cluster which might present the massive close-in exoplanets with
on-going tidal interaction in the ${\rm ln} P-{\rm ln} M$ space.
Our SCM found that there are four clusters in the $e$ space.
However, there is only one member in the Cluster $e_4$, which is
an unusual case with an extremely large eccentricity 0.927. The
other three clusters, Cluster $e_1$, $e_2$, and $e_3$ might
associate with the tidal, on-going tidal, and disc interactions,
respectively. The three clusters in the ${\rm ln} P-e$ space,
Cluster $Pe_1$, $Pe_2$, and $Pe_3$ could associate with the same
things, respectively. Finally, the eccentricity-period correlation is
strong and obvious.

\section*{Acknowledgment}
We would like to thank
the anonymous referee for their helpful comments and suggestions
to improve the presentation of the paper.
We are also grateful to the National Center for High-performance Computing
for computer time and facilities.
This work is supported in part
by the National Science Council, Taiwan, under
Ing-Guey Jiang's Grants: NSC 94-2112-M-008-010,
Li-Chin Yeh's Grants: NSC 94-2115-M-134-002,
Wen-Liang Hung's Grants: NSC 92-2213-E-134-001
and Miin-Shen Yang's Grants: NSC 93-2118-M-033-001.

\clearpage

\begin{figure}
\epsscale{.50}
 \plotone{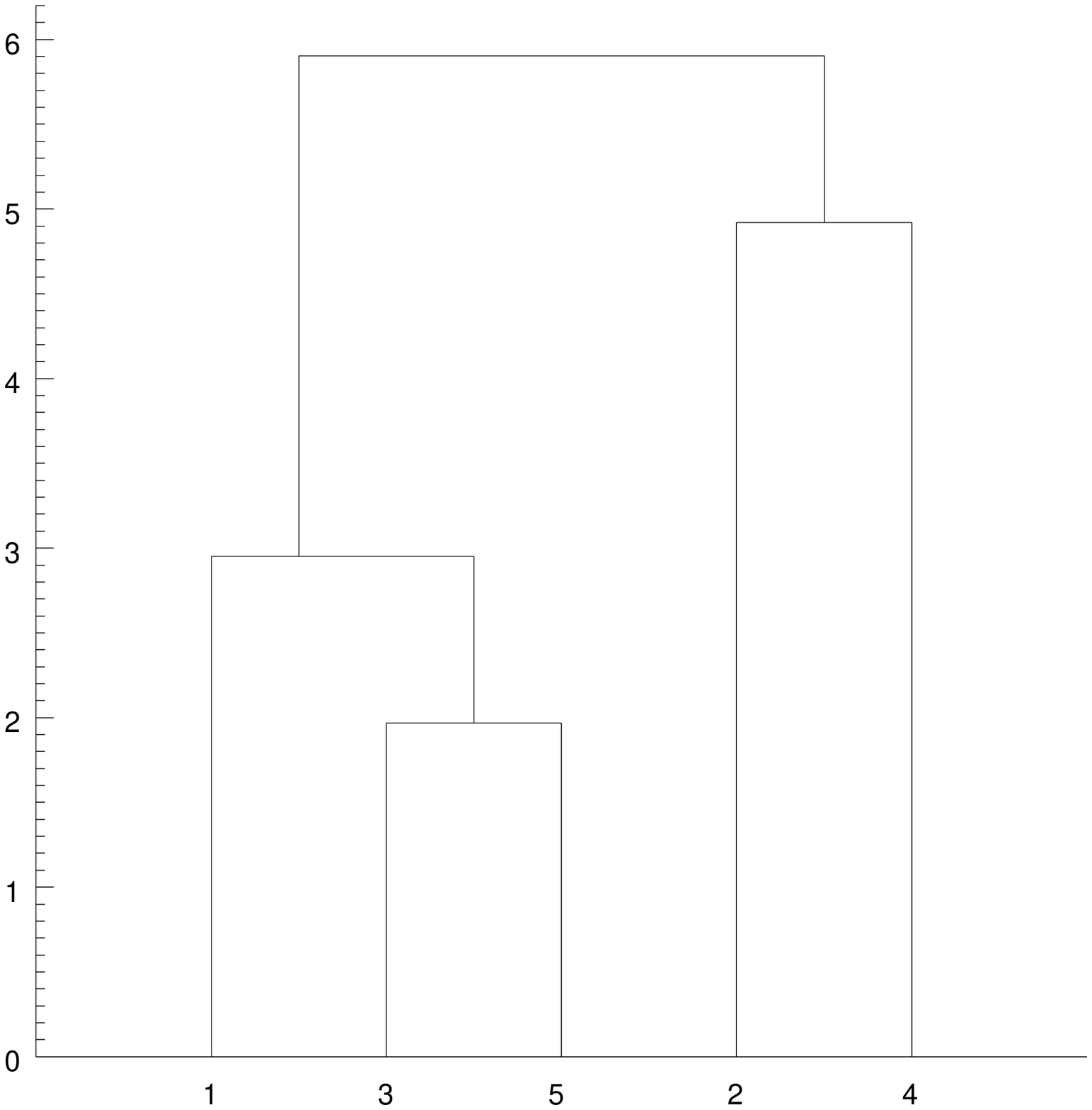}
  \caption{The dendrogram for distances between five objects in the example
of Section 2.}
\end{figure}

\clearpage

\begin{figure}
\epsscale{.50}
 \plotone{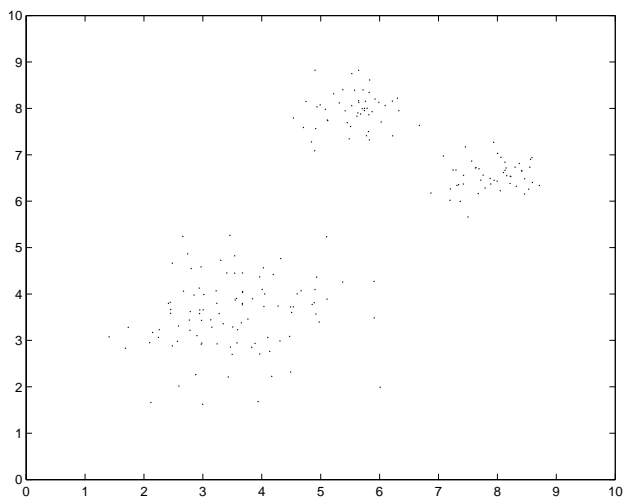}
  \caption{The data set with one larger cluster and two small clusters.}
\end{figure}

\begin{figure}
\epsscale{1.0}
 \plottwo{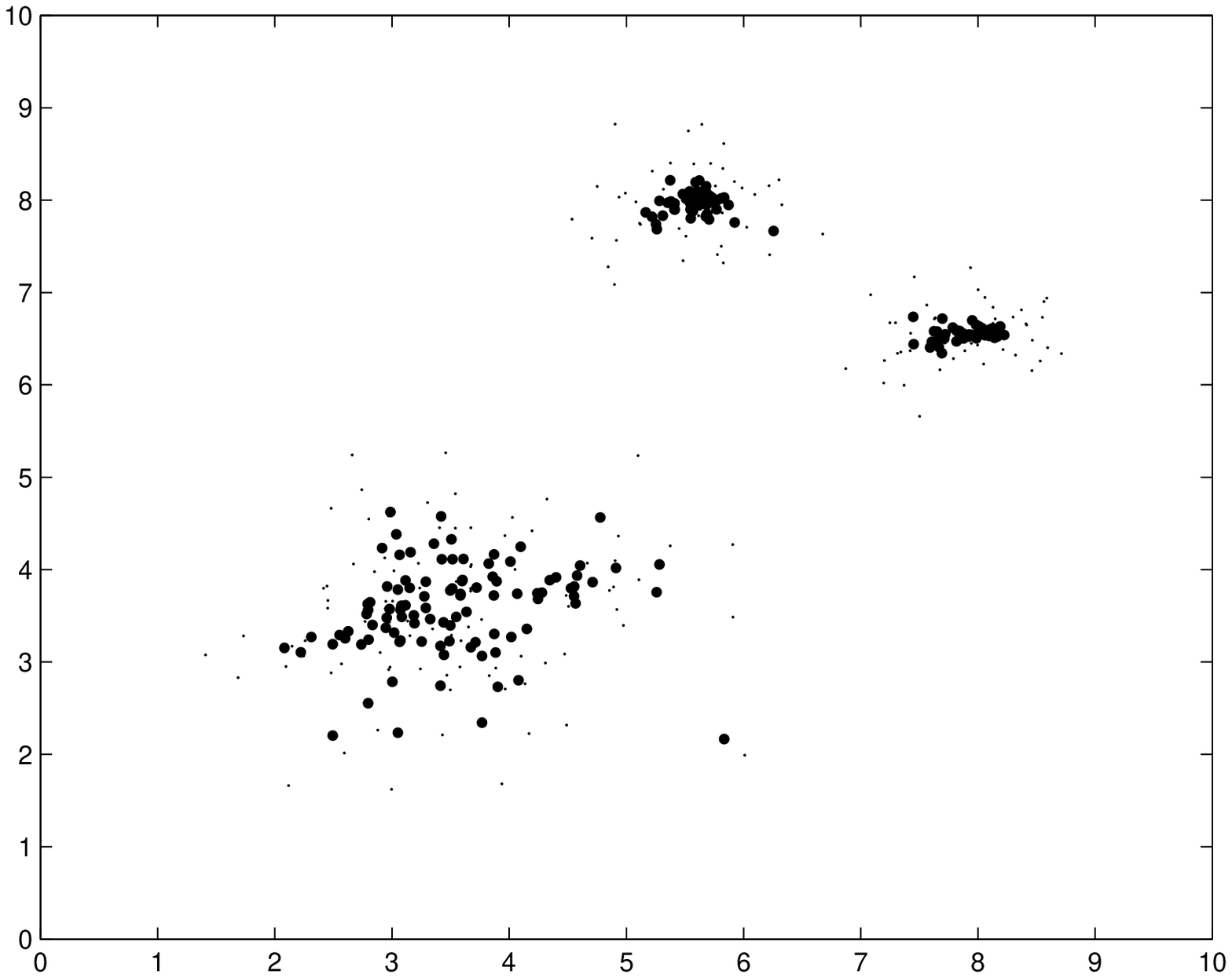}{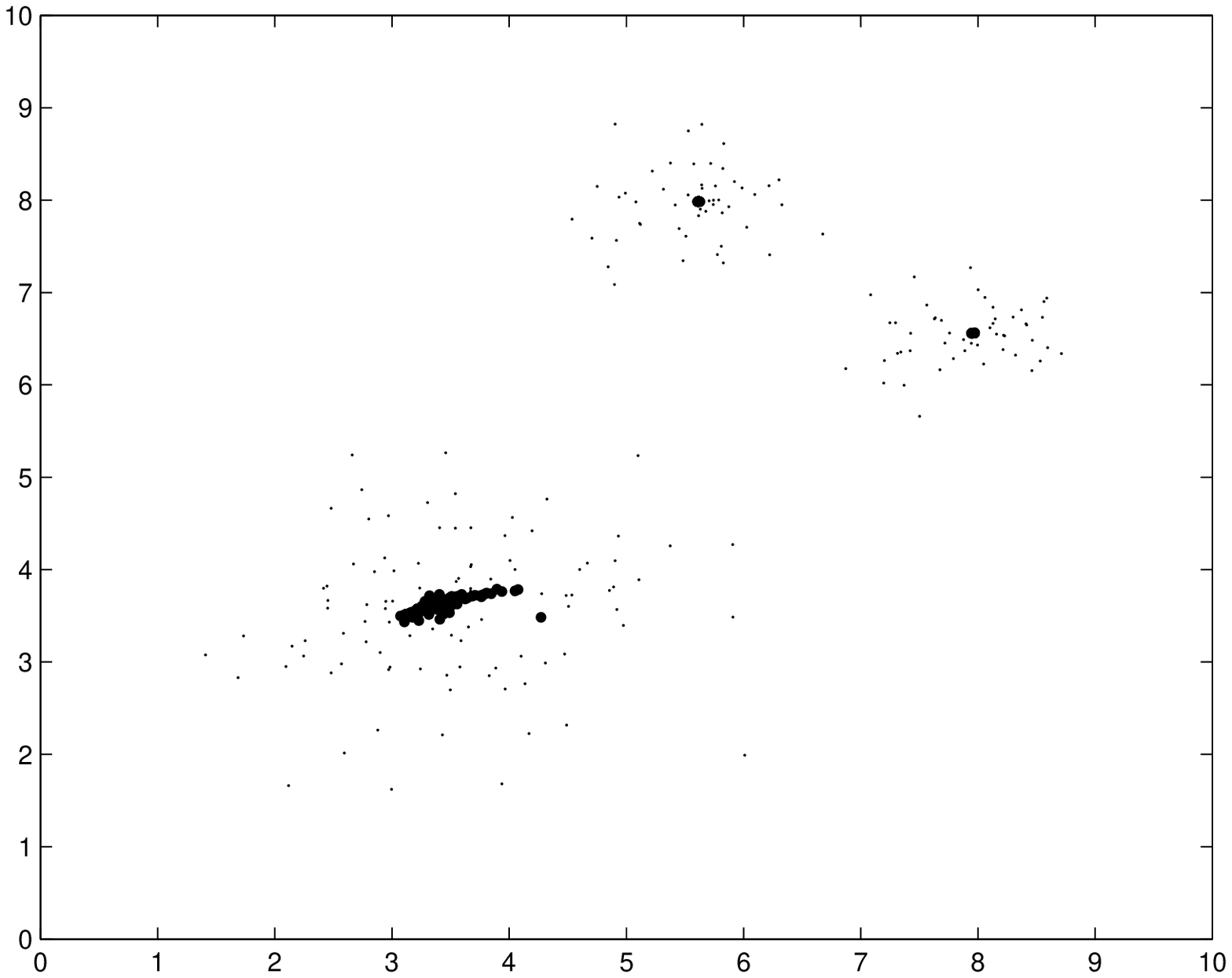}
  \caption{The positions of cluster centers (full circles):
the left panel is after 1 iteration and the
 right panel is after 5 iterations. Please note that the original data points
are indicated by dots.}
\end{figure}

\clearpage
\begin{figure}
\epsscale{1.0}
 \plottwo{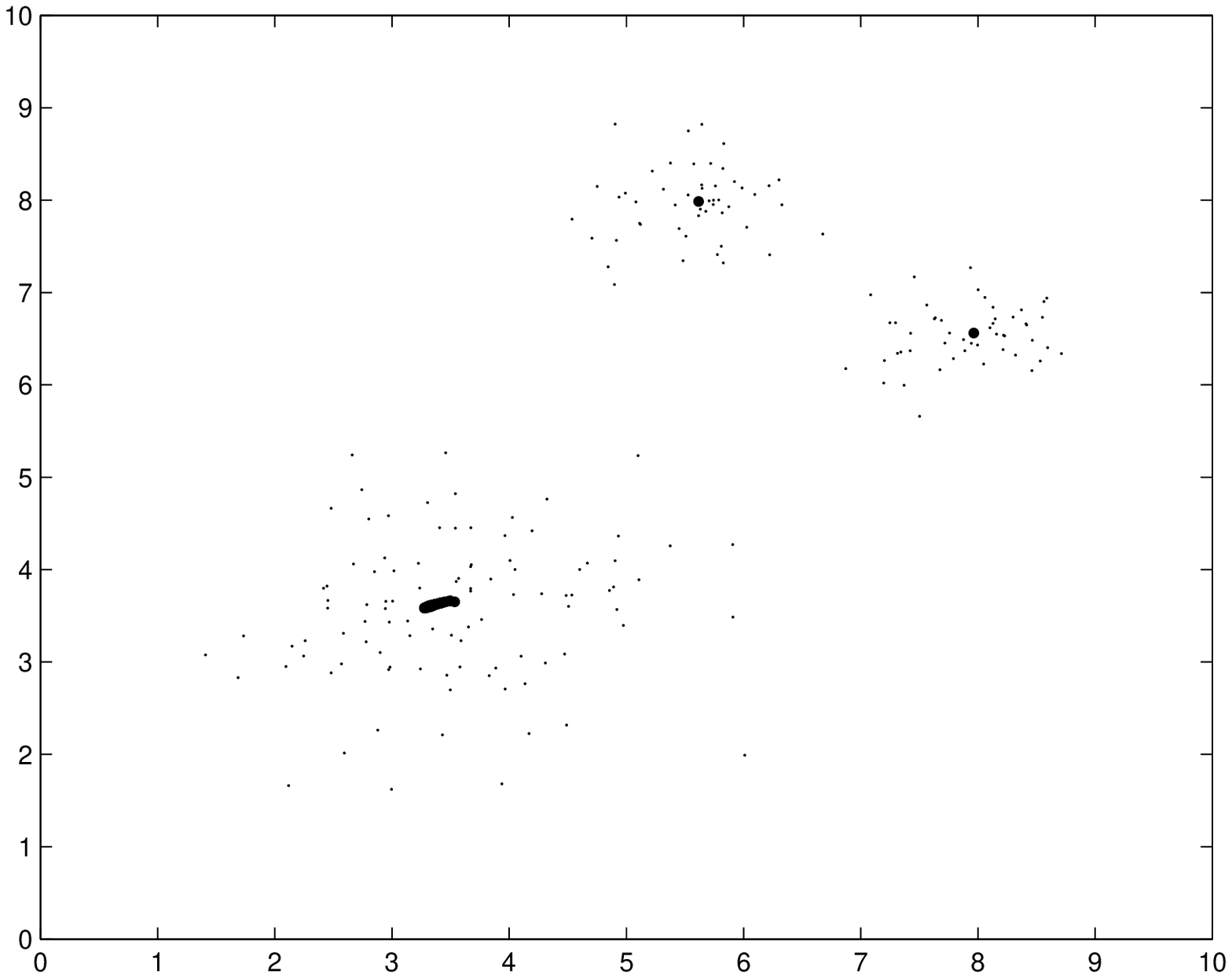}{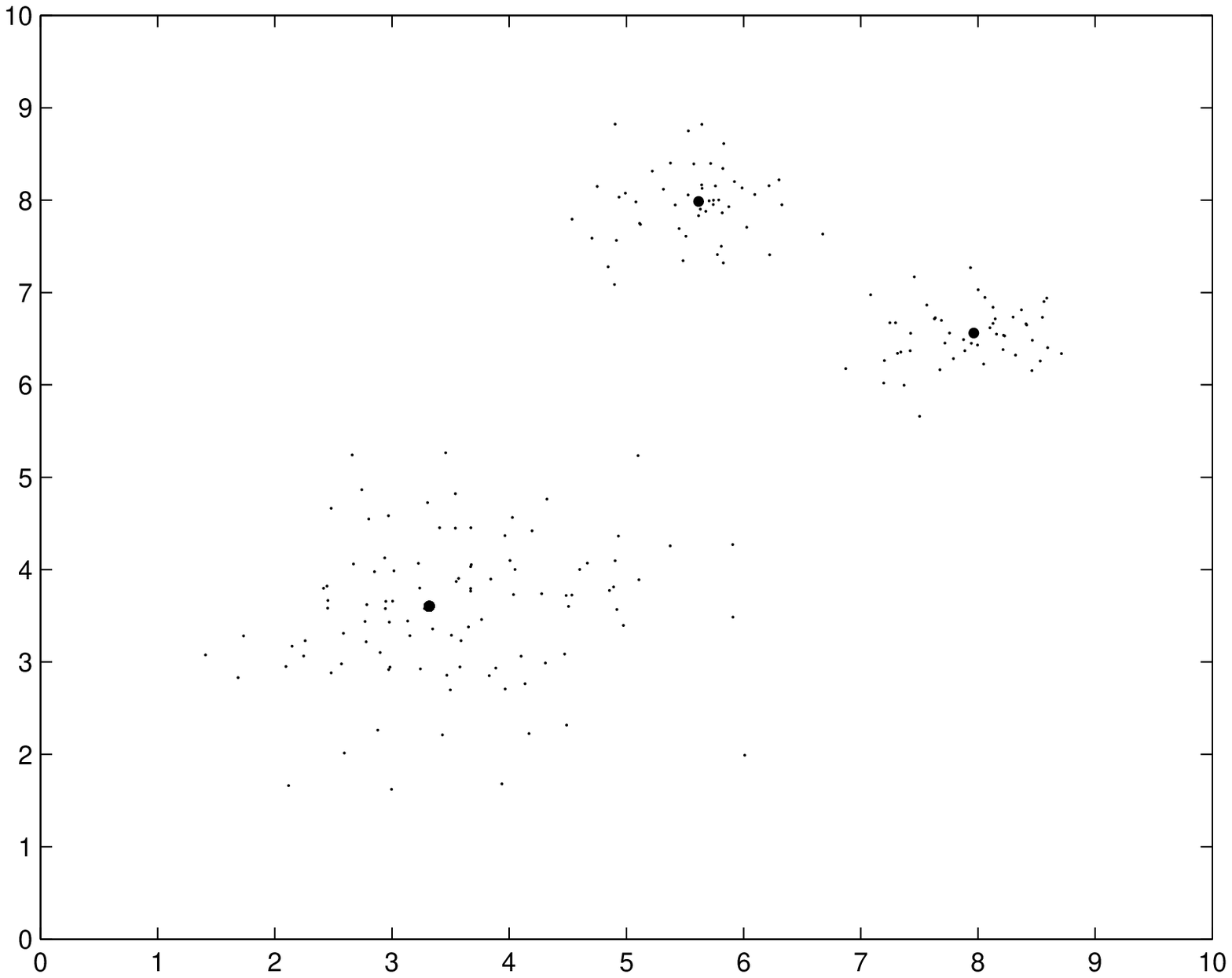}
  \caption{The positions of cluster centers (full circles):
the left panel is after 10 iteration and the
 right panel is the convergent. Please note that the original data points
are indicated by dots.}
\end{figure}

\begin{figure}
\epsscale{1.0}
 \plottwo{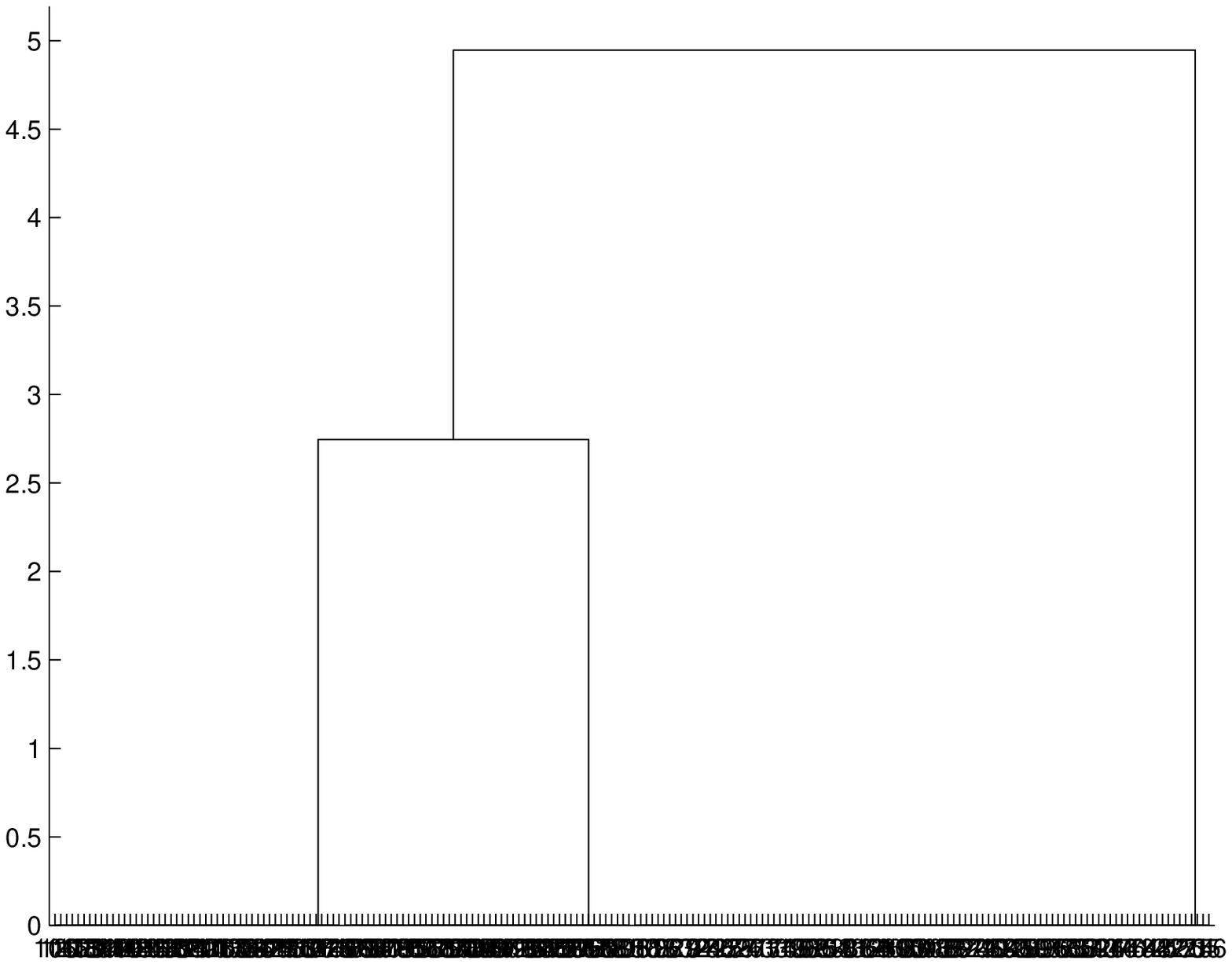}{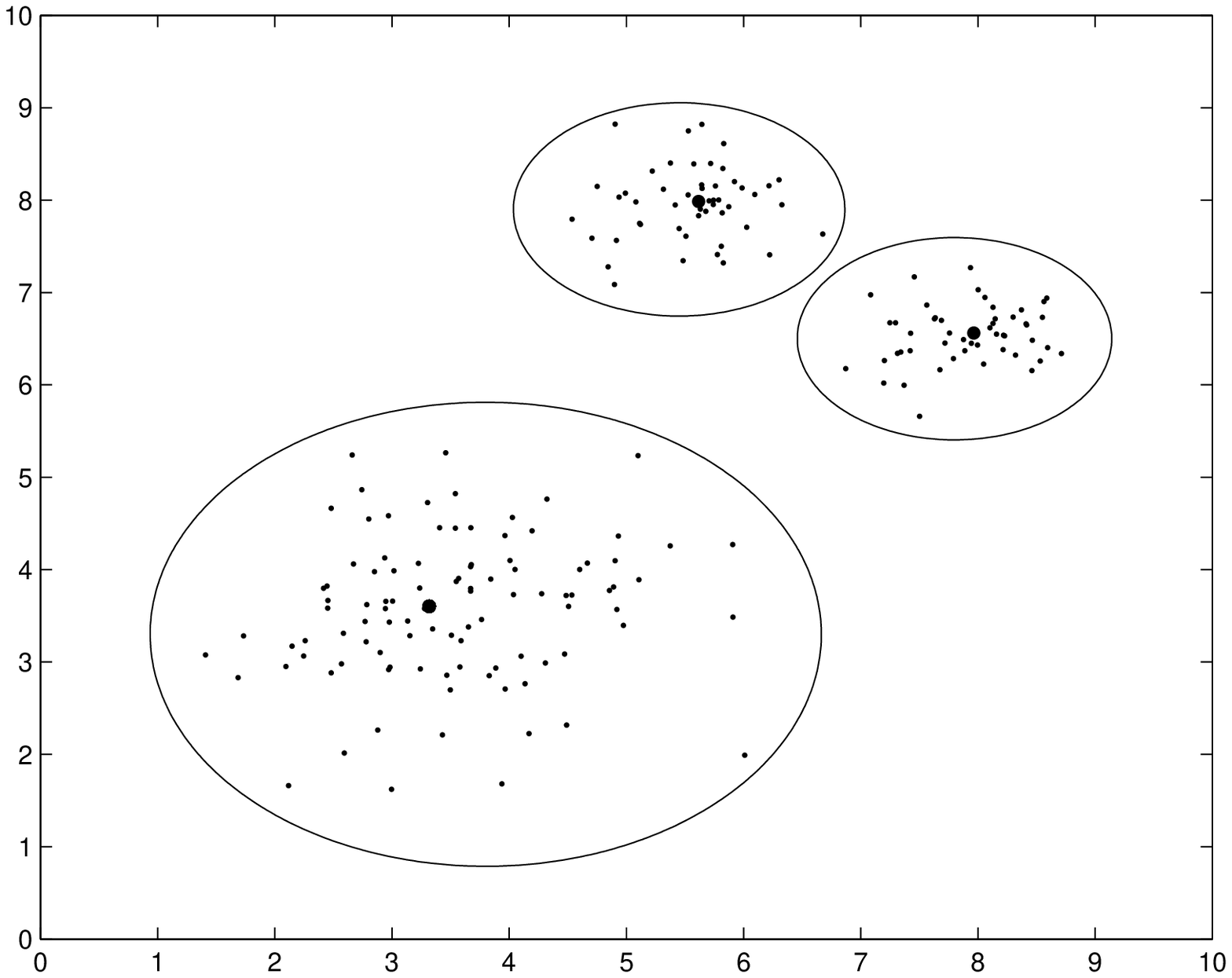}
  \caption{The left panel is the dendrogram through the SCM.
There are too many data points, so the 
 data identity numbers below the horizontal axis are not clear.
The right panel
is the identified clusters through the SCM.}
\end{figure}

\clearpage

\begin{figure}
\epsscale{.50}
 \plotone{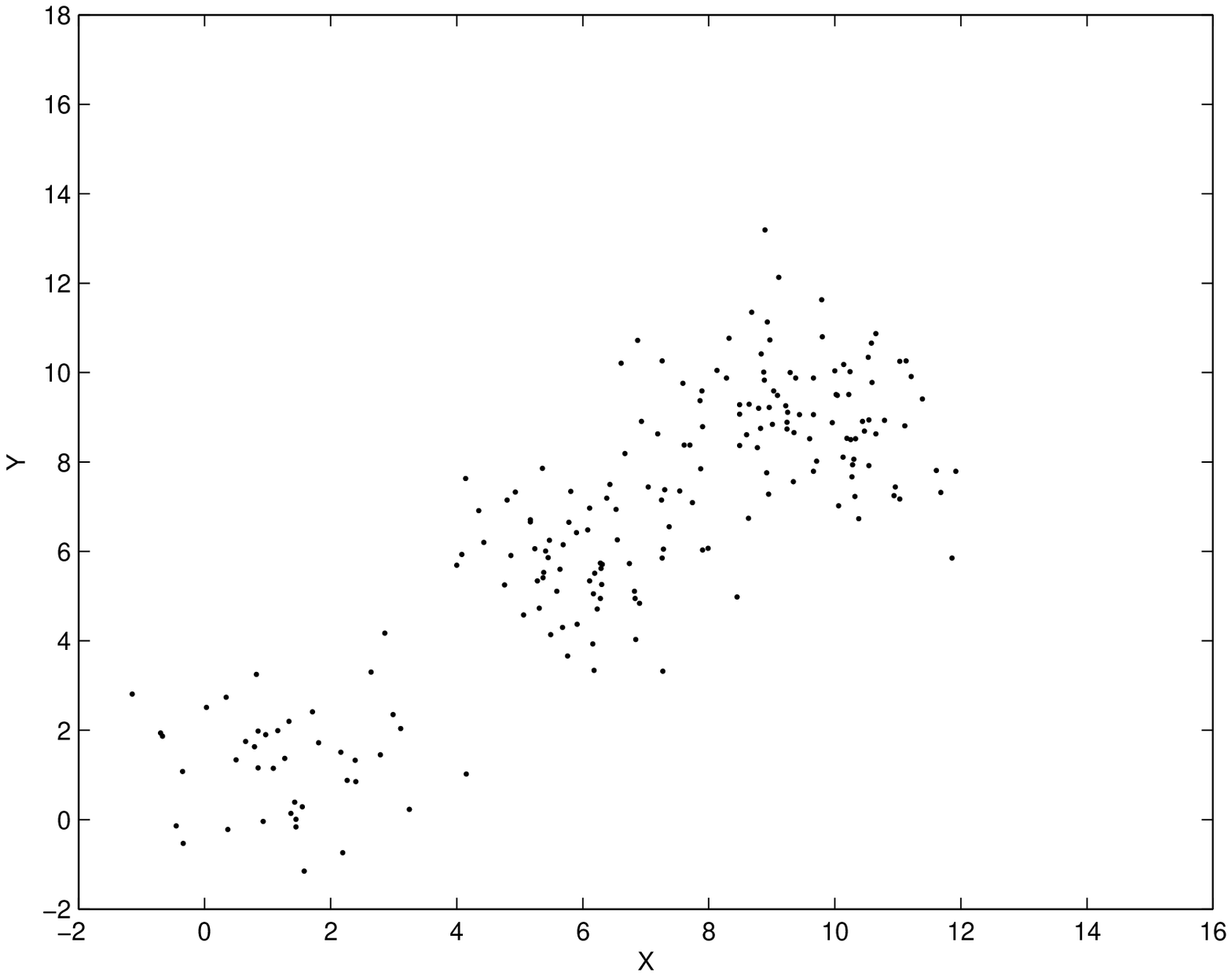}
  \caption{The data set generated for Test 1 }
\end{figure}

\begin{figure}
\epsscale{1.0}
 \plottwo{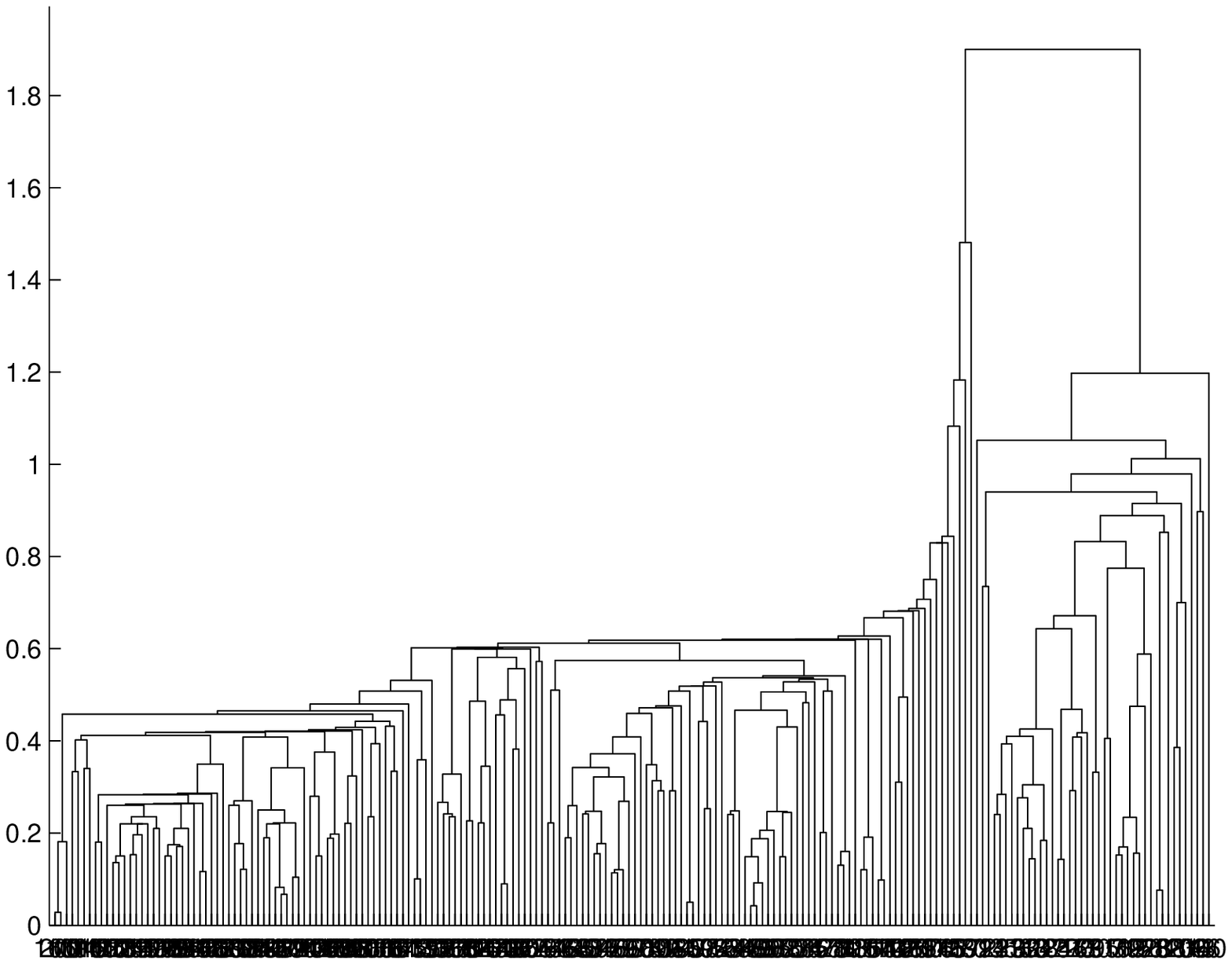}{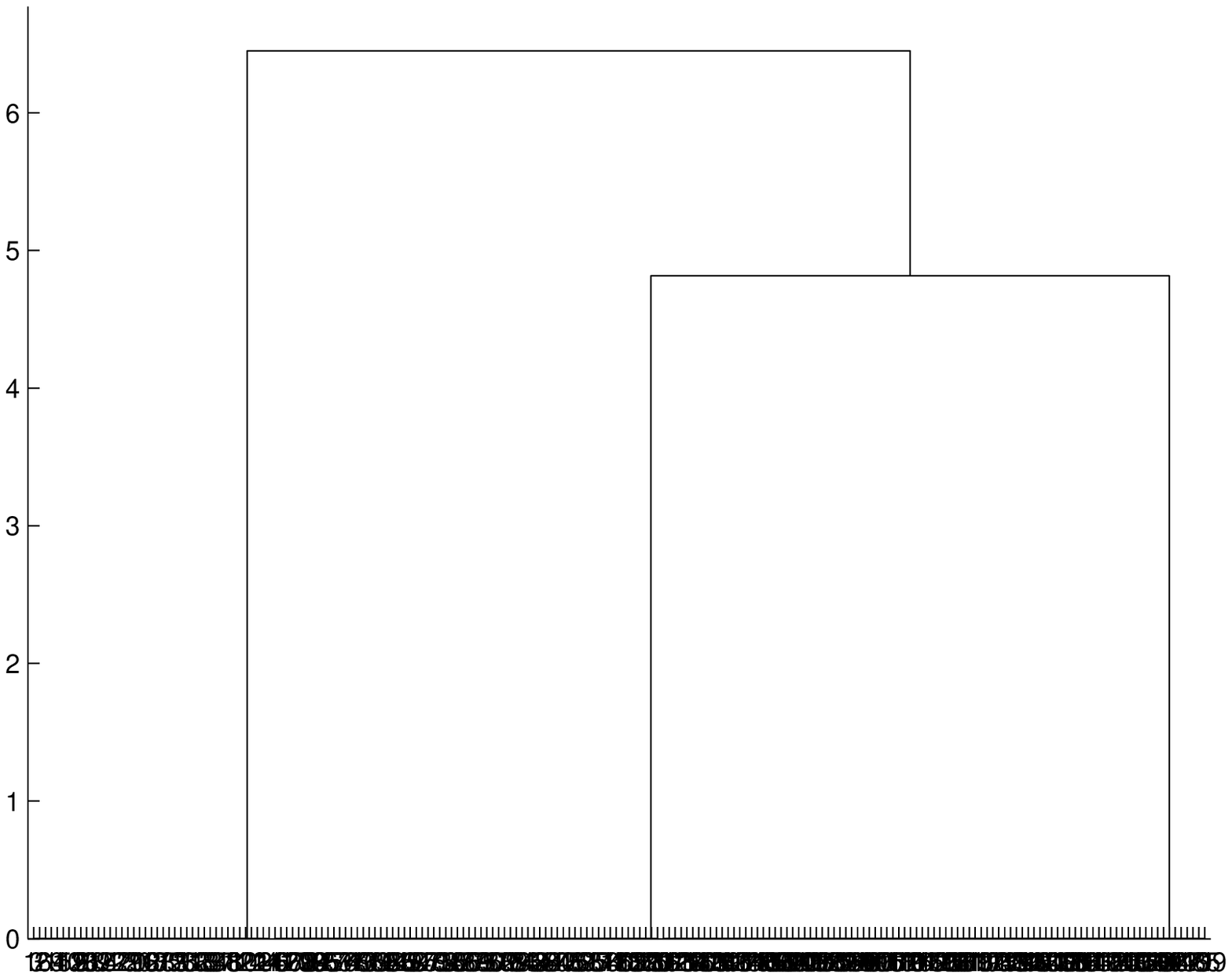}
  \caption{The dendrogram of Test 1:
the single linkage algorithm's result is on the left and the SCM's
is on the right.
There are too many data points, so the 
 data identity numbers below the horizontal axis are not clear.}
\end{figure}

\clearpage

\begin{figure}
\epsscale{.50}
 \plotone{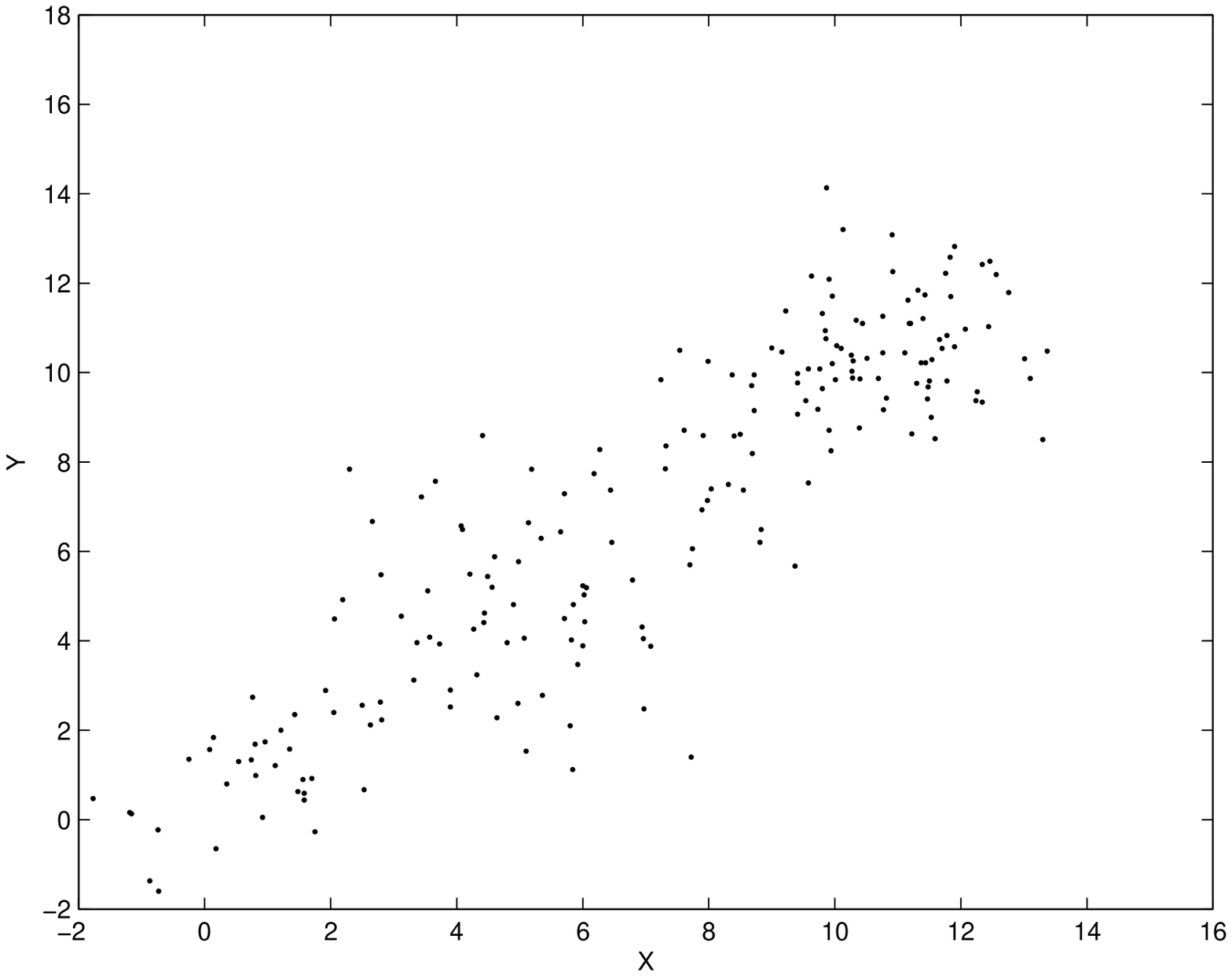}
  \caption{The data set generated for Test 2 }
\end{figure}

\begin{figure}
\epsscale{1.0}
 \plottwo{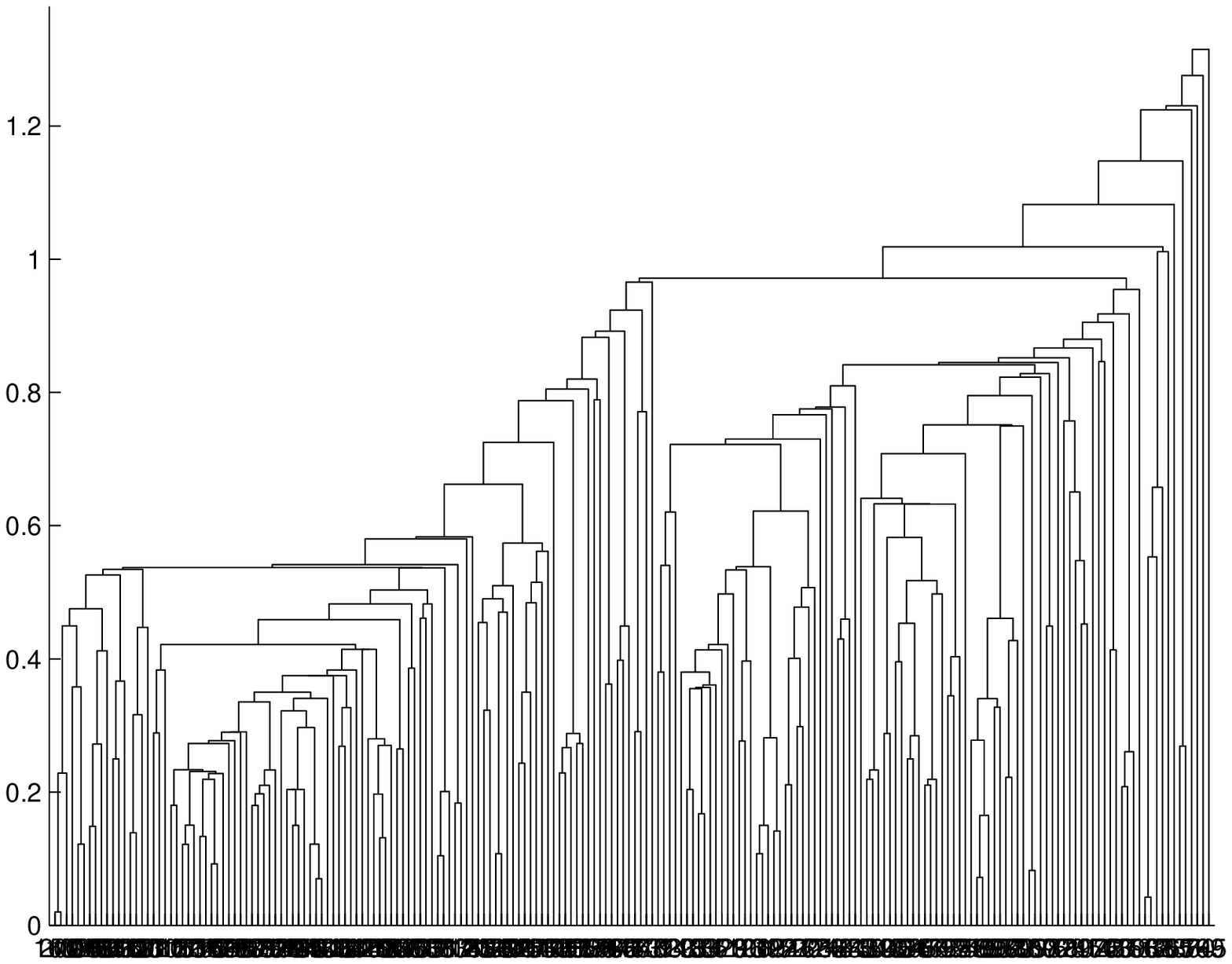}{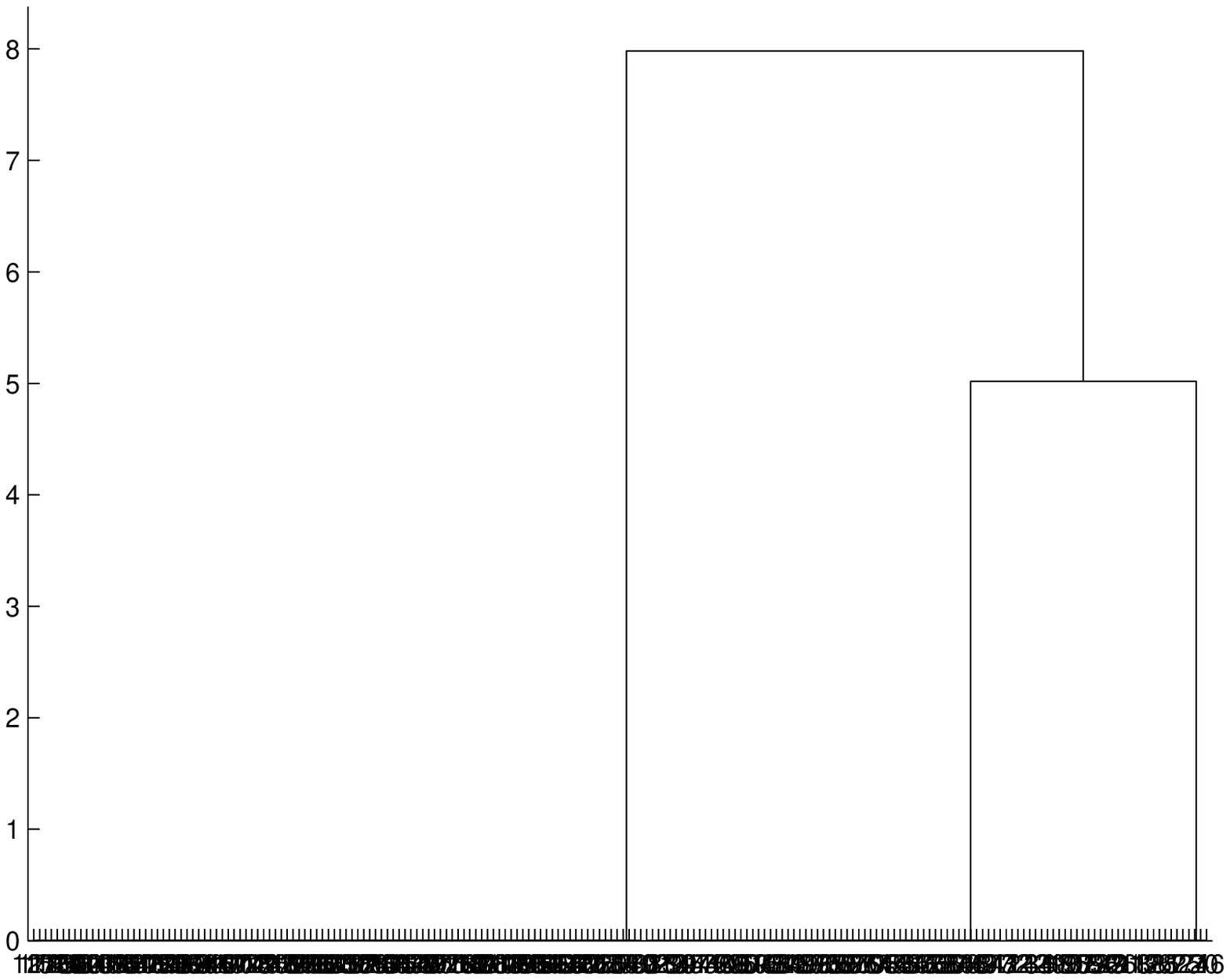}
  \caption{
The dendrogram of Test 2:
the single linkage algorithm's result is on the left and the SCM's
is on the right.
There are too many data points, so the 
 data identity numbers below the horizontal axis are not clear.
}
\end{figure}

\clearpage

\begin{figure}
\epsscale{.50}
 \plotone{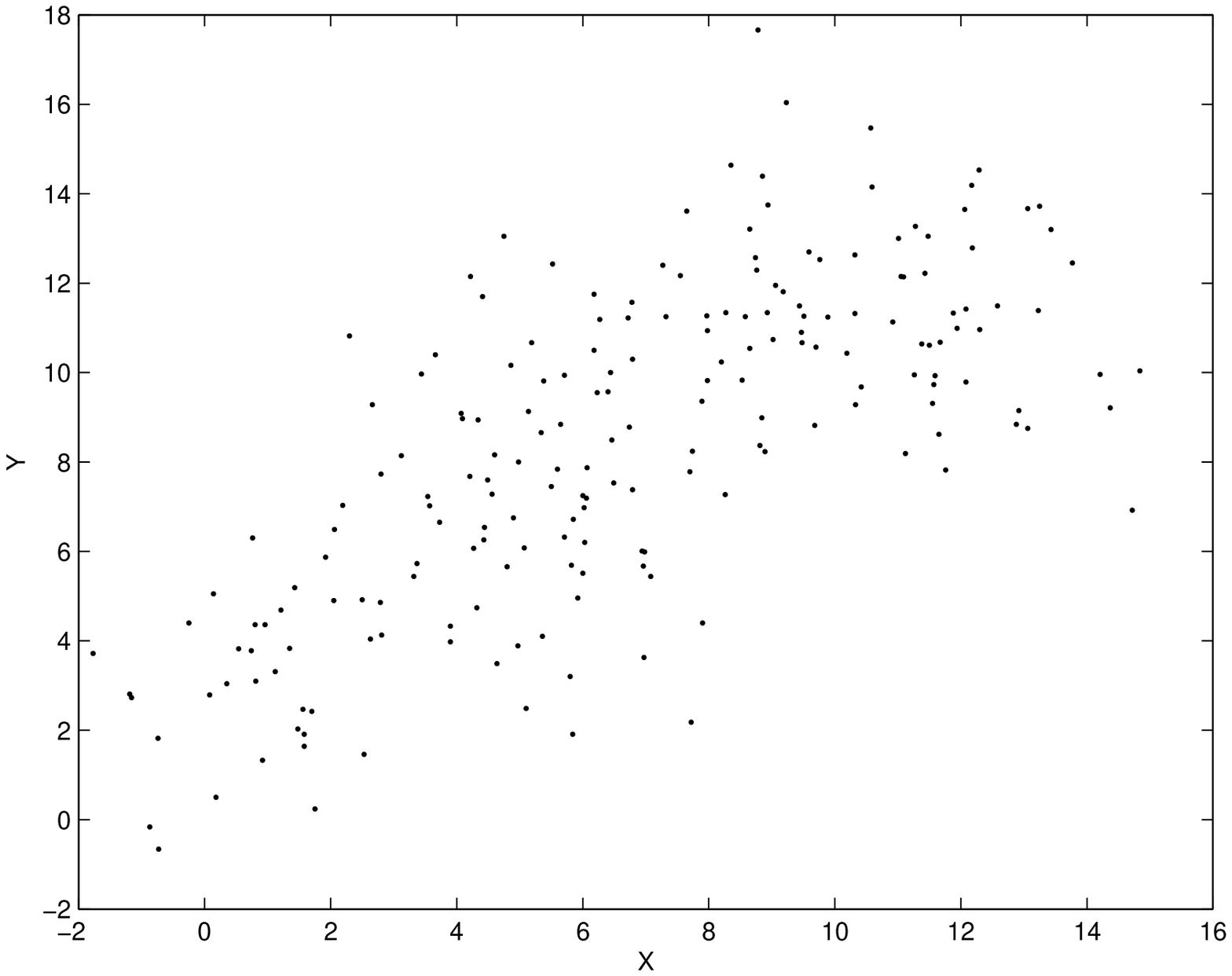}
  \caption{The data set generated for Test 3 }
\end{figure}

\begin{figure}
\epsscale{1.0}
 \plottwo{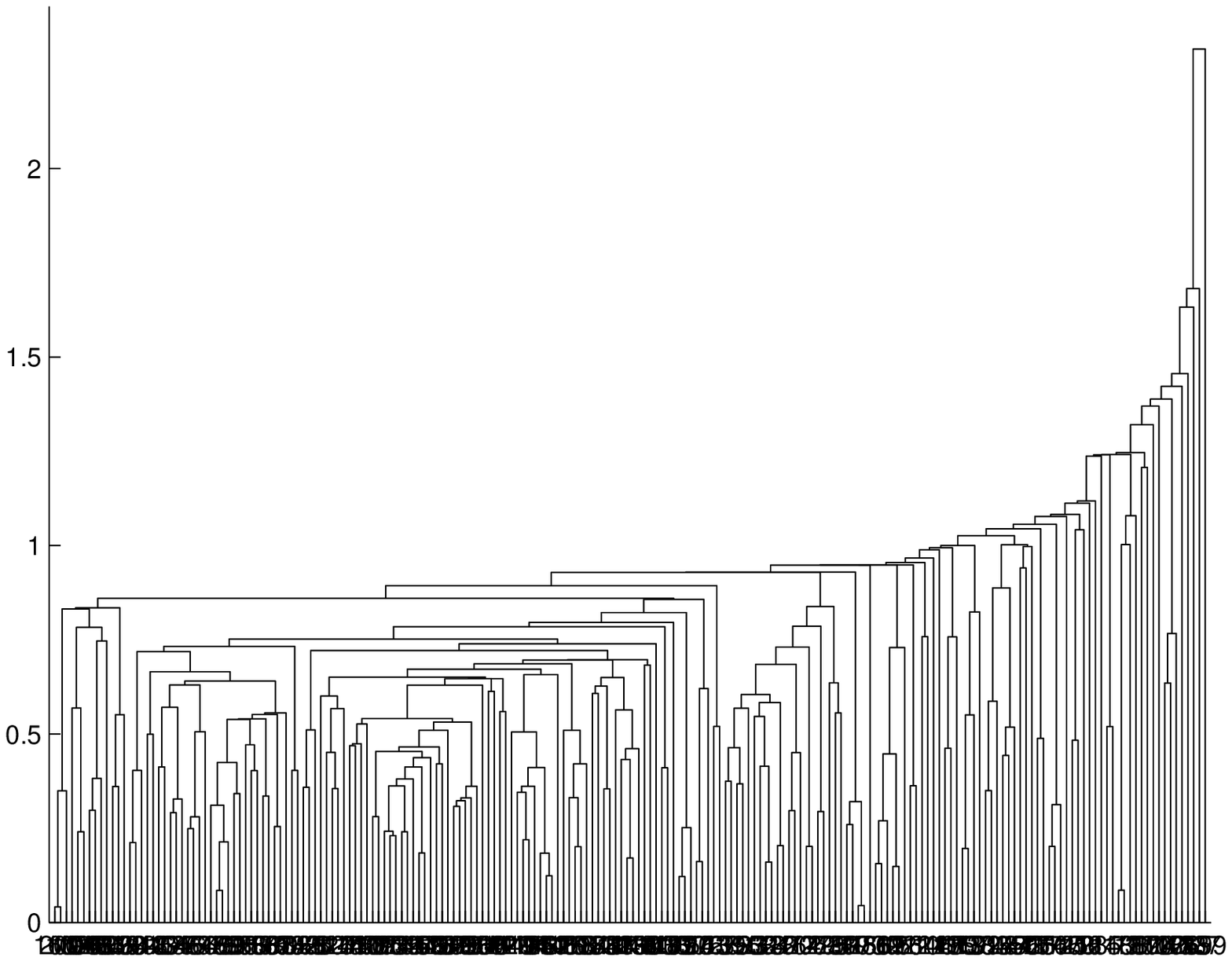}{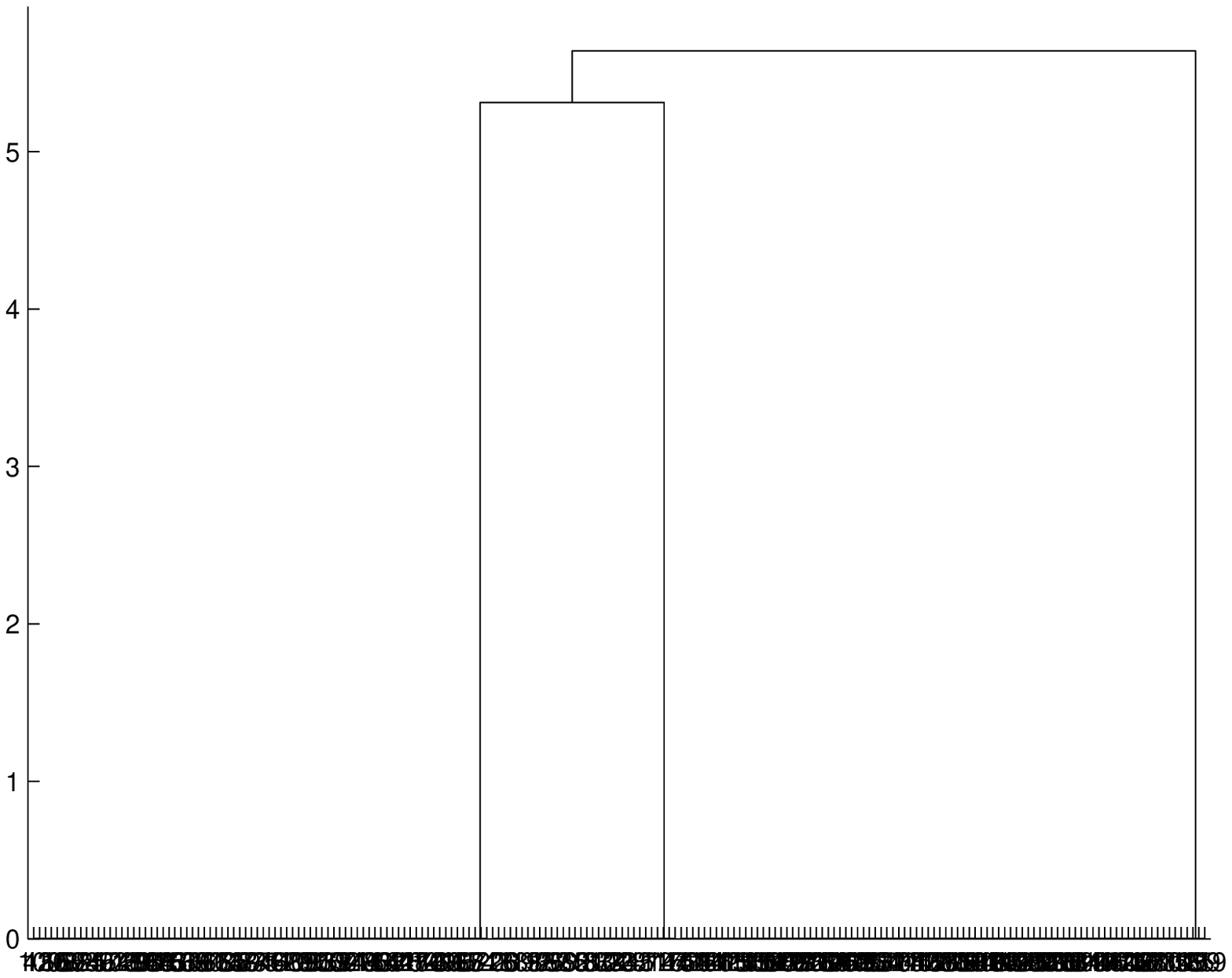}
  \caption{
The dendrogram of Test 3:
the single linkage algorithm's result is on the left and the SCM's
is on the right.
There are too many data points, so the 
 data identity numbers below the horizontal axis are not clear.
}
\end{figure}

\clearpage

\begin{figure}
\epsscale{1.0} \plottwo{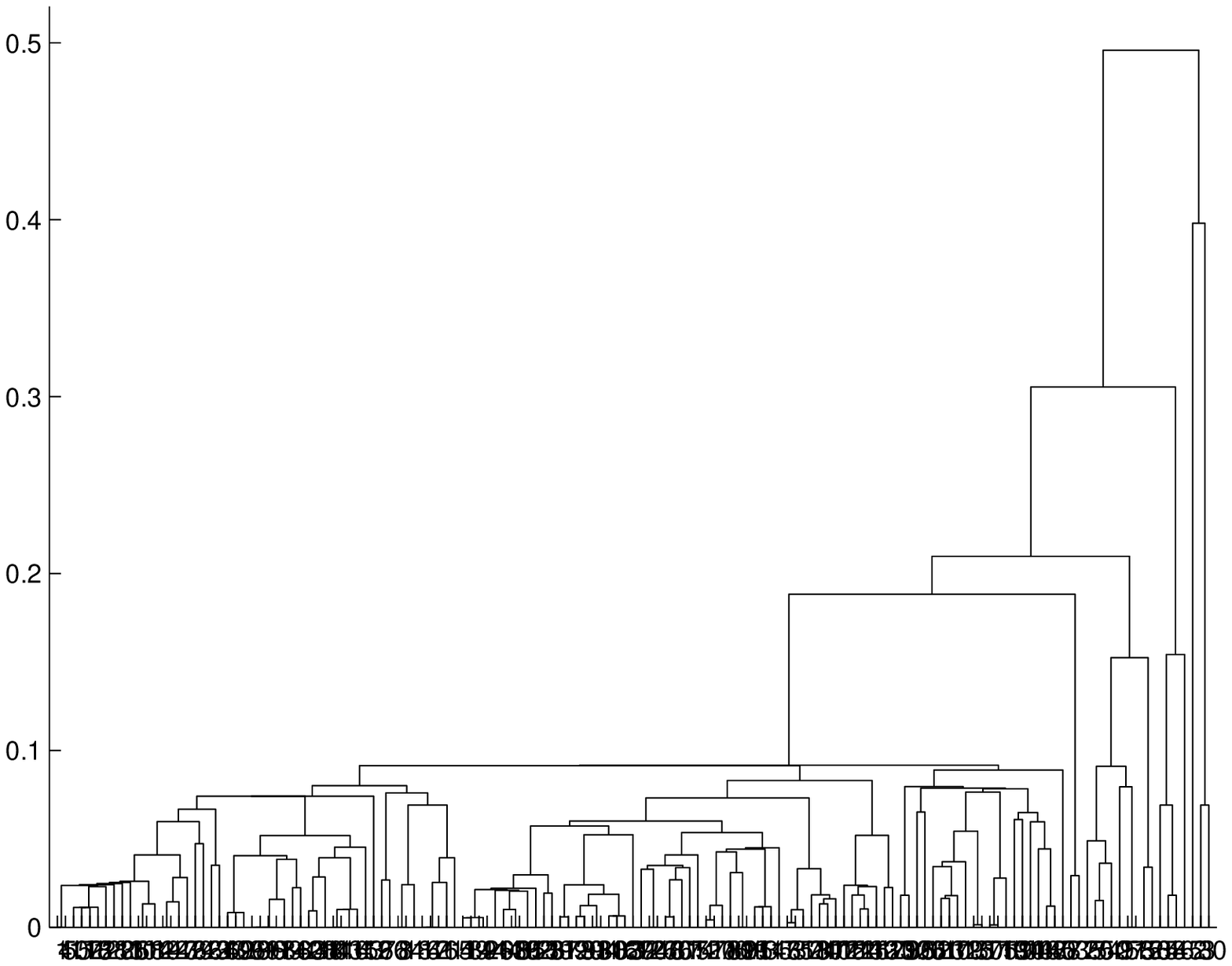}{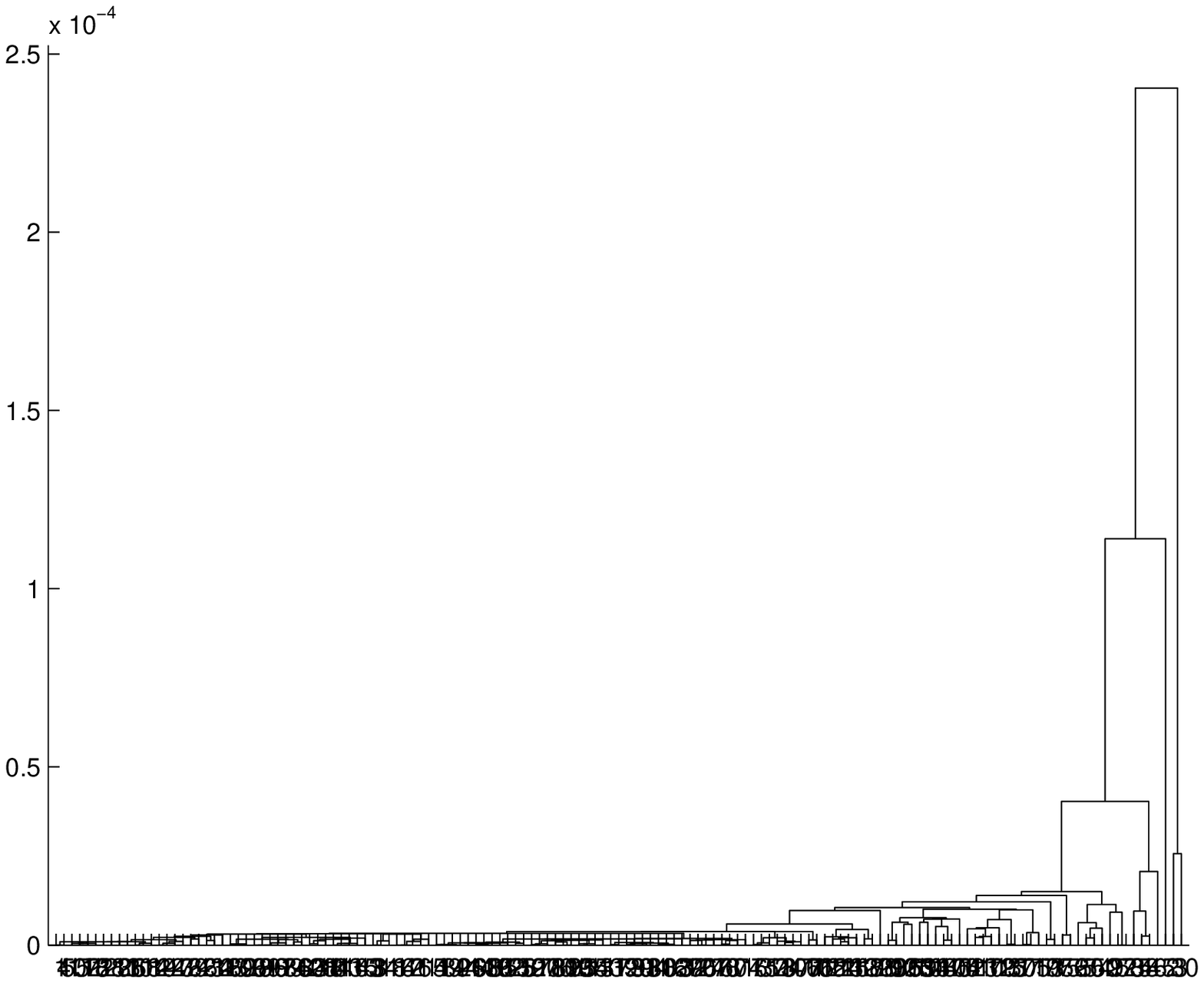}
  \caption{The dendrogram of exoplanets in the ${\rm ln} M$ space: the
single linkage algorithm's result is on the left and the SCM's is
on the right.
 There are too many data points, so the 
 data identity numbers below the horizontal axis are not clear.
}
\end{figure}

\begin{figure}
\epsscale{.50} \plotone{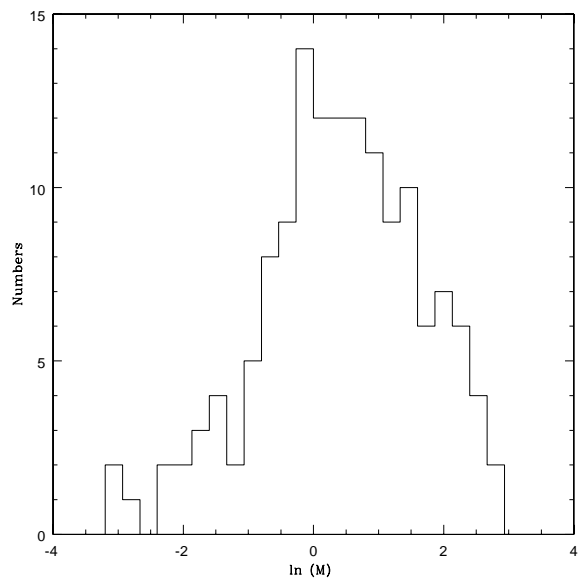}
 \caption{The histogram of exoplanets in the ${\rm ln} M$ space.}
\end{figure}

\clearpage

\begin{figure}
\epsscale{1.0} \plottwo{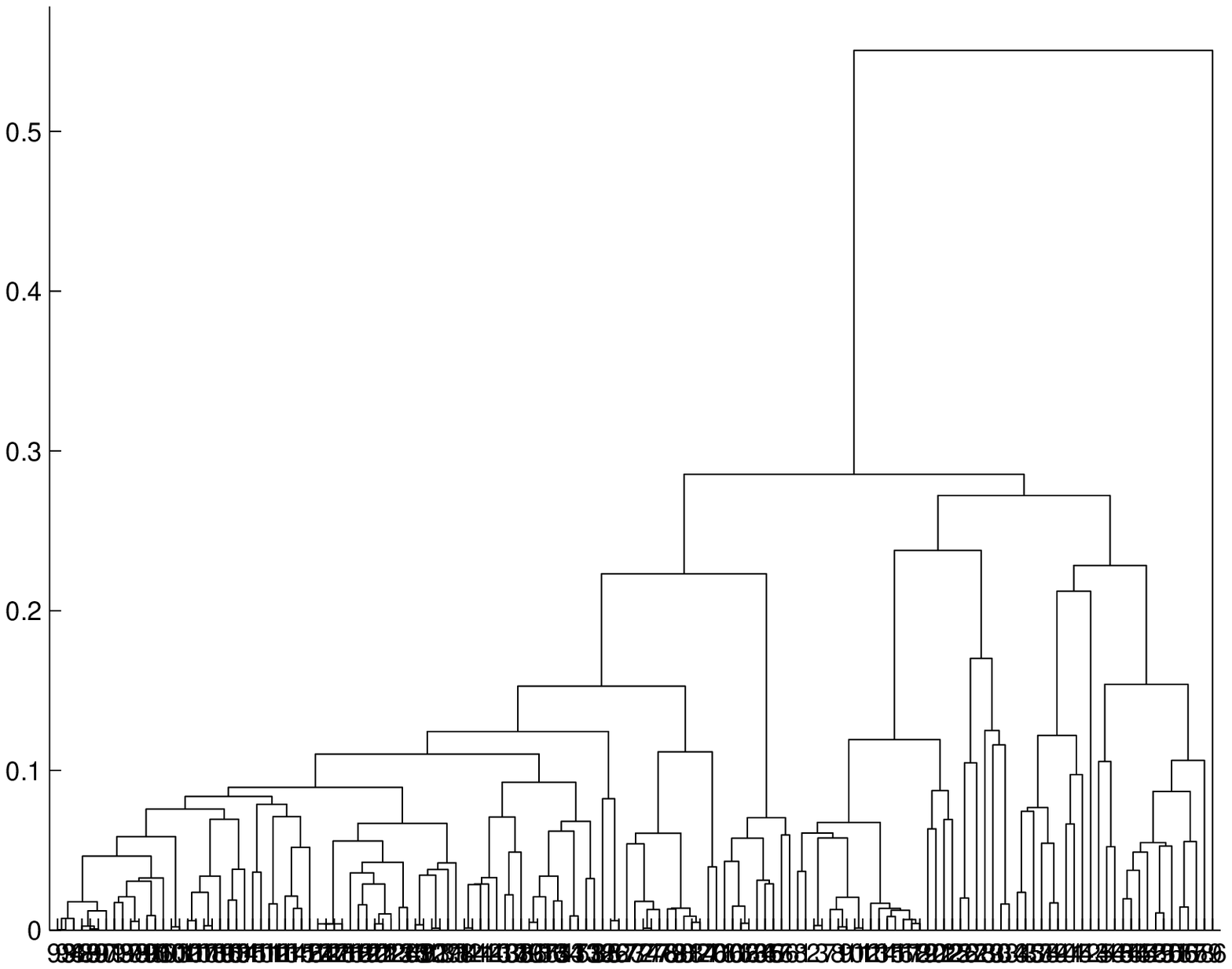}{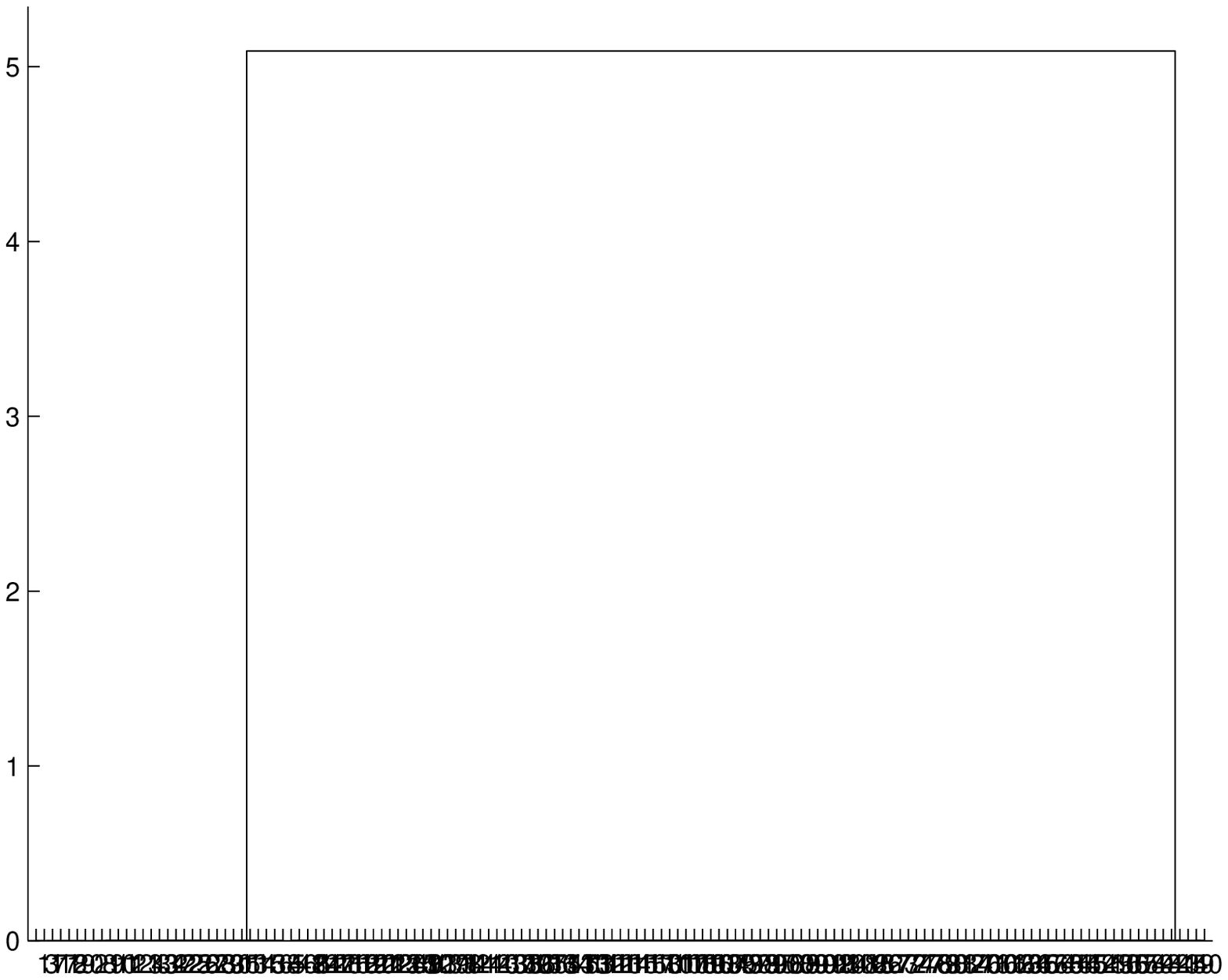}
\caption{The dendrogram of exoplanets in the ${\rm ln} P$ space: the single
linkage algorithm's result is on the left and the SCM's is on the
right. 
There are too many data points, so the 
 data identity numbers below the horizontal axis are not clear.
}
\end{figure}

\begin{figure}
\epsscale{.50}
 \plotone{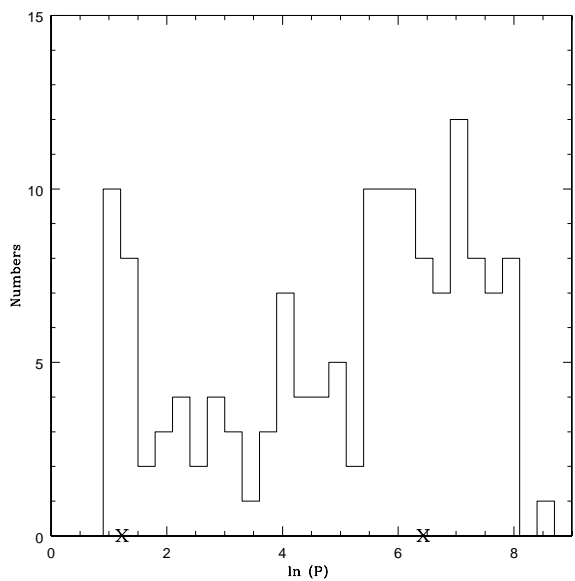}
 \caption{The histogram of exoplanets in the ${\rm ln} P$ space. In the plot,
the left cross indicates that the center of Cluster $P_1$ is at
1.4552 and the right cross indicates that the center of Cluster
$P_2$ is at 6.5455.}
\end{figure}

\clearpage

\begin{figure}
\epsscale{1} \plottwo{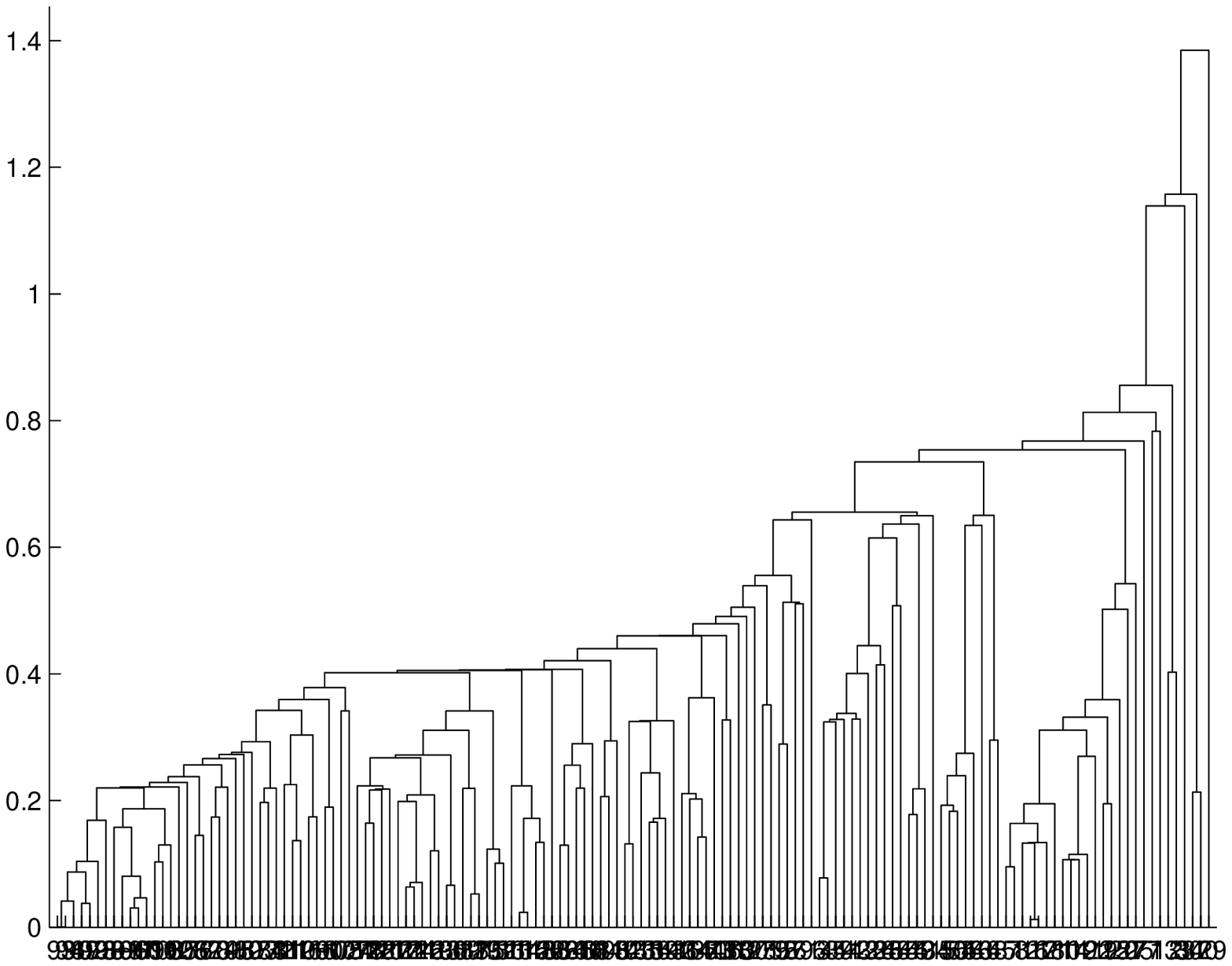}{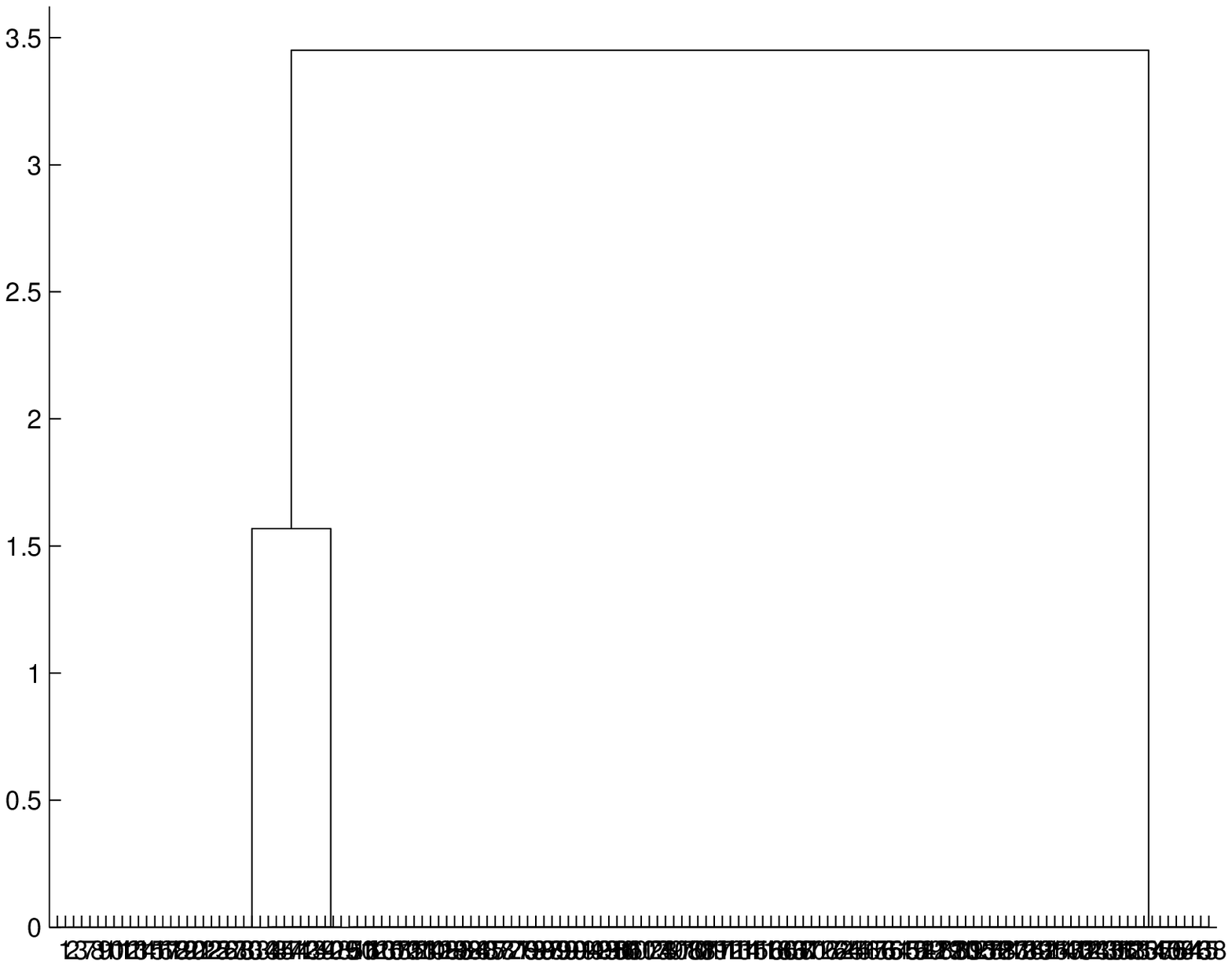}
 \caption{The dendrogram of exoplanets in the ${\rm ln} P-{\rm ln} M$ space: 
the single linkage algorithm's result is on the left and the SCM's is
on the right. 
There are too many data points, so the 
 data identity numbers below the horizontal axis are not clear.
}
\end{figure}

\begin{figure}
\epsscale{0.6}
\plotone{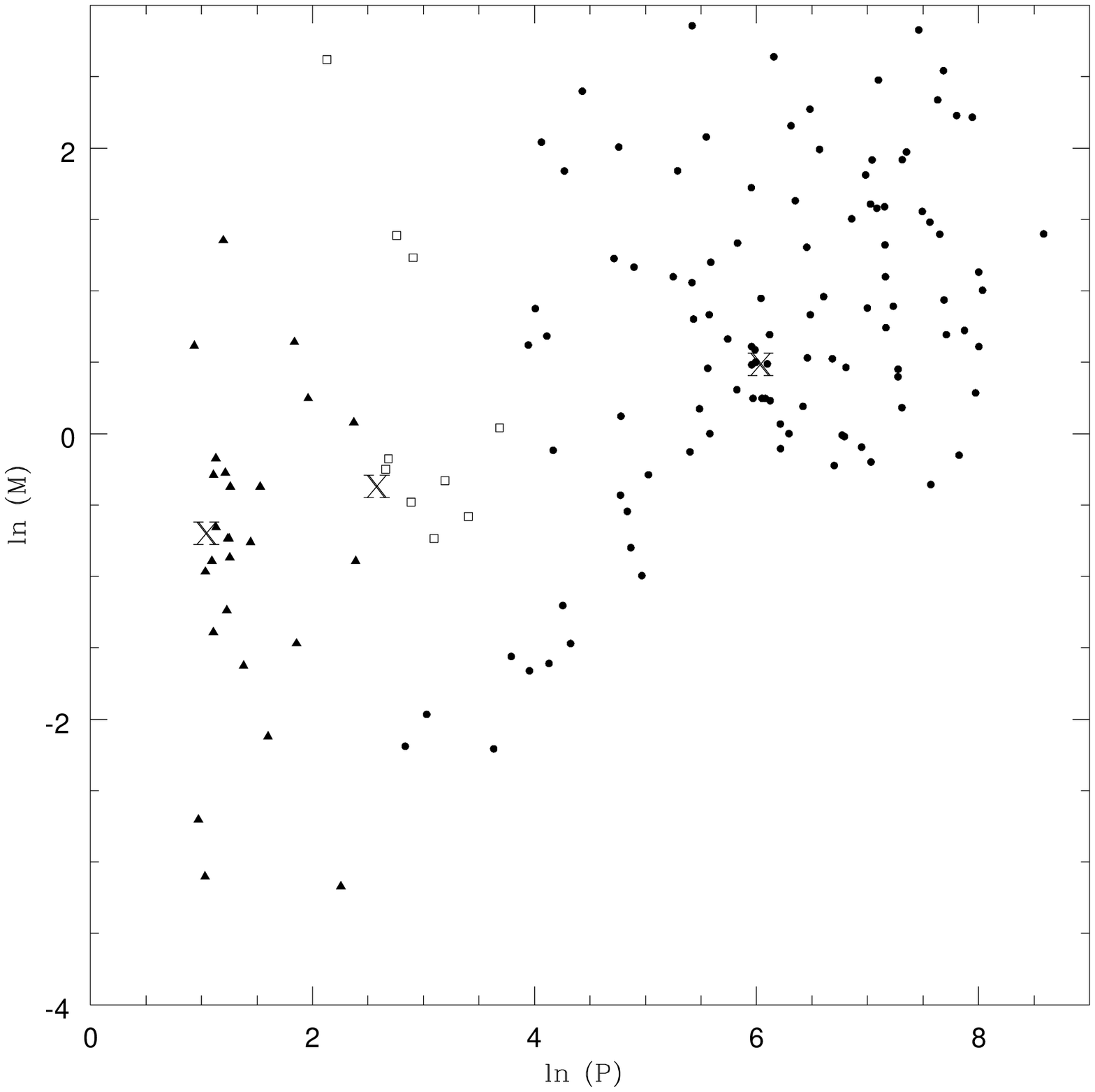}
 \caption{The distribution of exoplanets in the ${\rm ln} P-{\rm ln} M$ space.
There are three crosses indicating the centers of three clusters.
The triangles indicate the members of the Cluster $PM_1$, the open
squares indicate the members of the Cluster $PM_2$, and the full
circles indicate the members of the Cluster $PM_3$.}
\end{figure}

\clearpage
\begin{figure}
\epsscale{1.0}
\plottwo{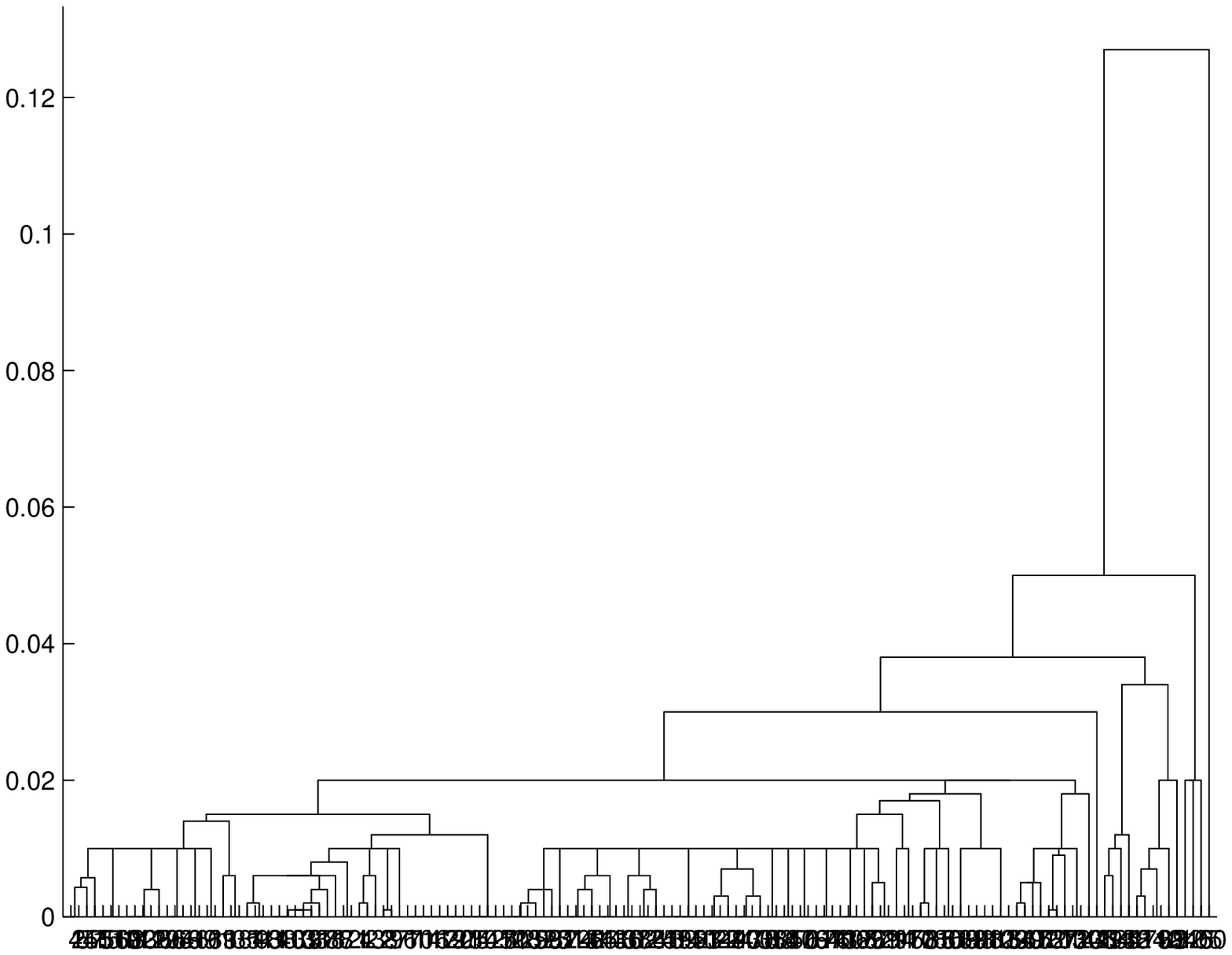}{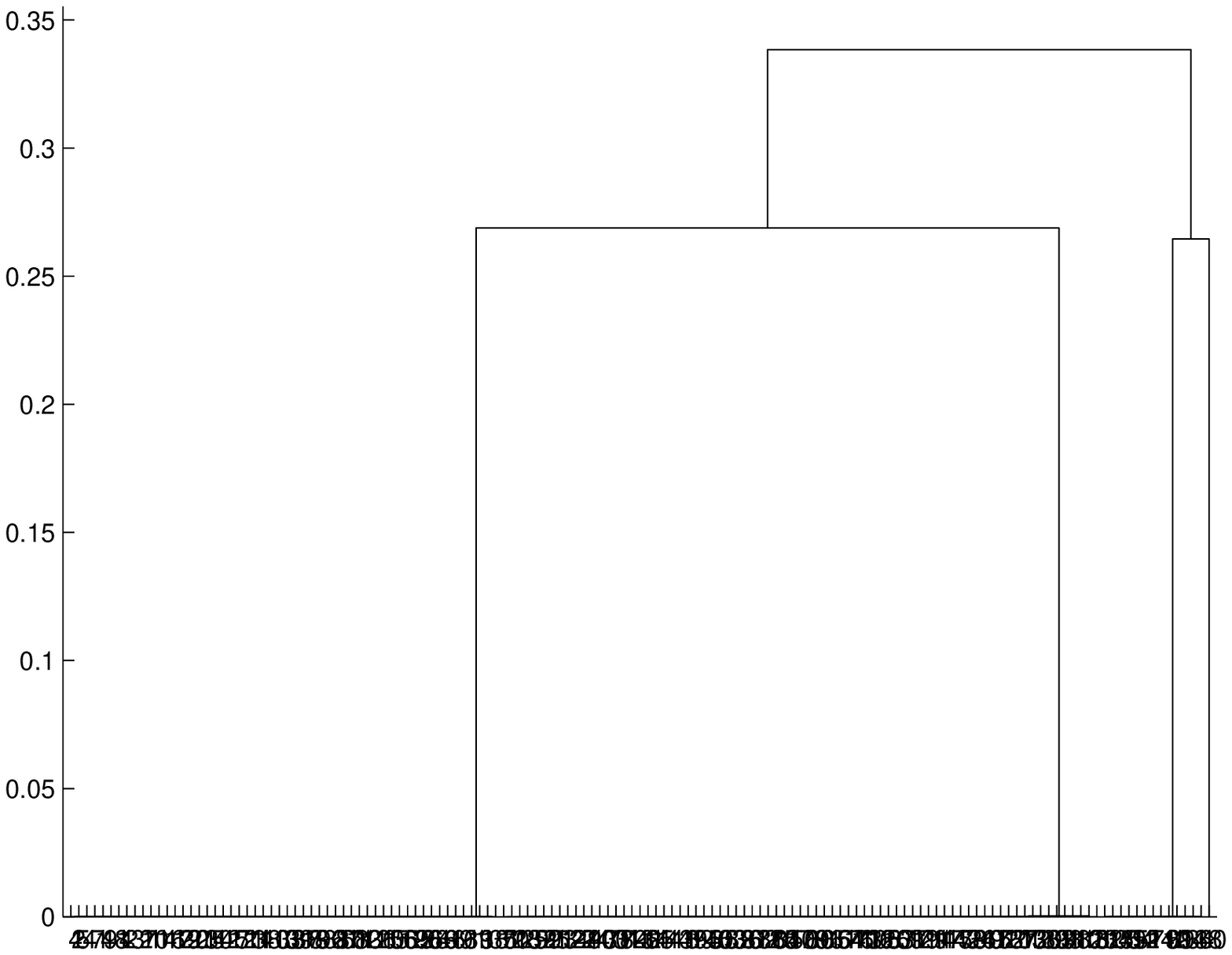}
 \caption{The dendrogram of exoplanets in the $e$ space: the single linkage
algorithm's result is on the left and the SCM's is on the right.
There are too many data points, so the 
 data identity numbers below the horizontal axis are not clear.
 }
\end{figure}

\begin{figure}
\epsscale{.50} \plotone{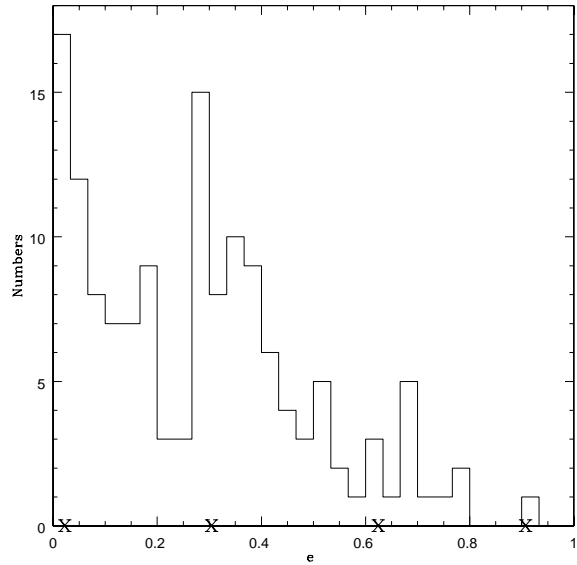}
 \caption{The histogram of exoplanets in the $e$ space. There are four crosses
indicating the centers of four clusters.}
\end{figure}

\clearpage
\begin{figure}
\epsscale{1.0} \plottwo{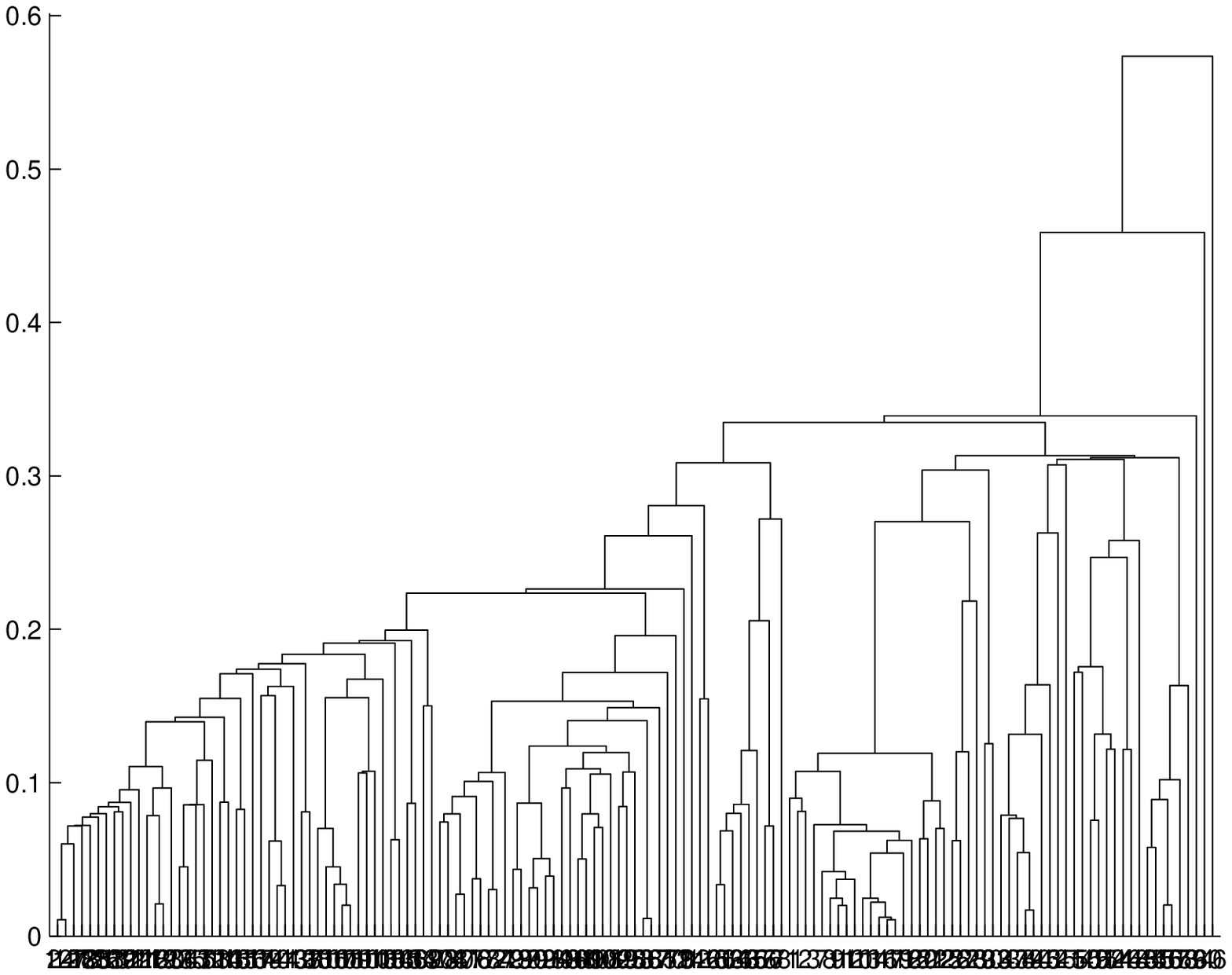}{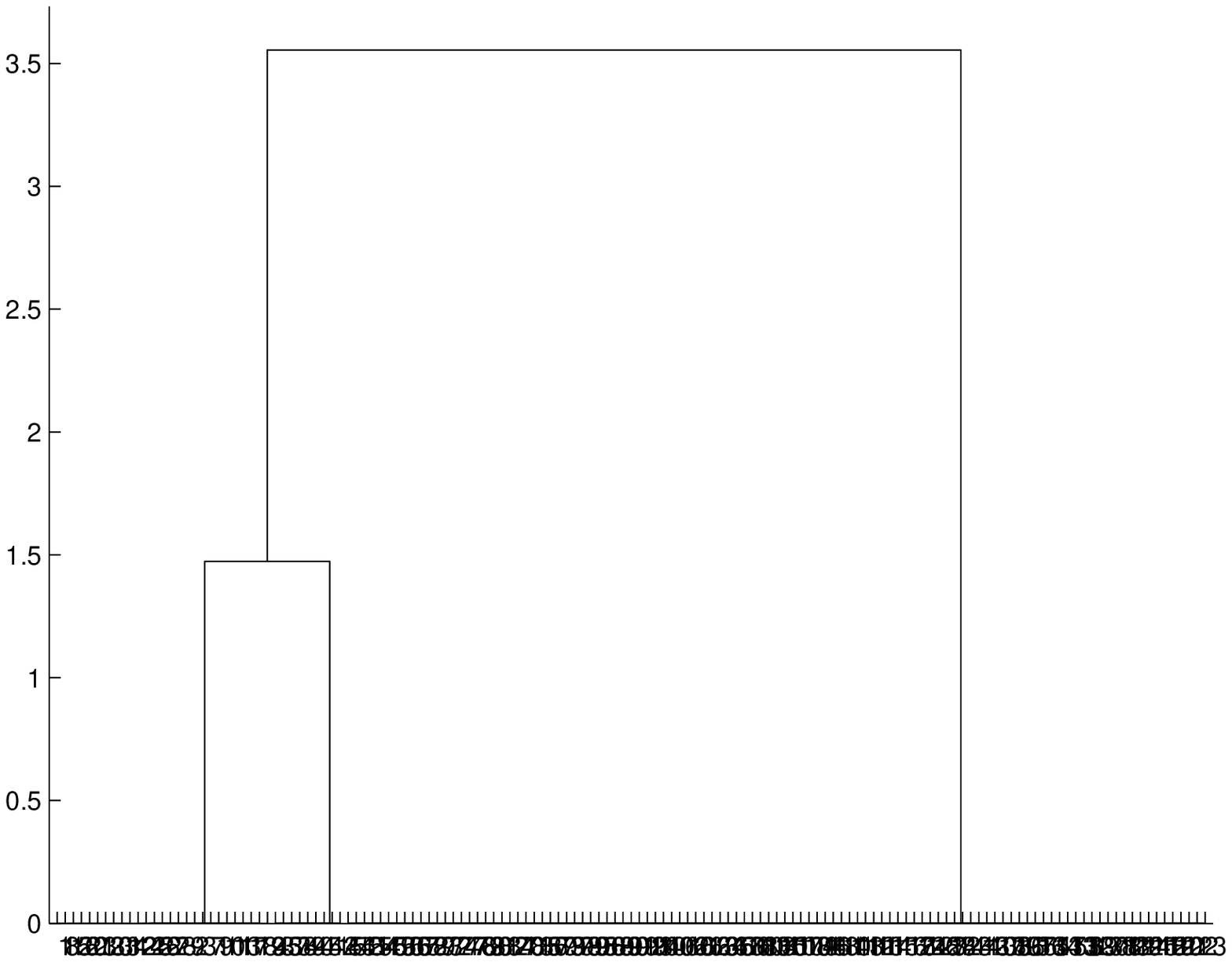}
 \caption{The dendrogram of exoplanets in the ${\rm ln} P-e$ space: the single
linkage algorithm's result is on the left and the SCM's is on the
right. 
There are too many data points, so the 
 data identity numbers below the horizontal axis are not clear.
}
\end{figure}

\begin{figure}
\epsscale{0.6} \plotone{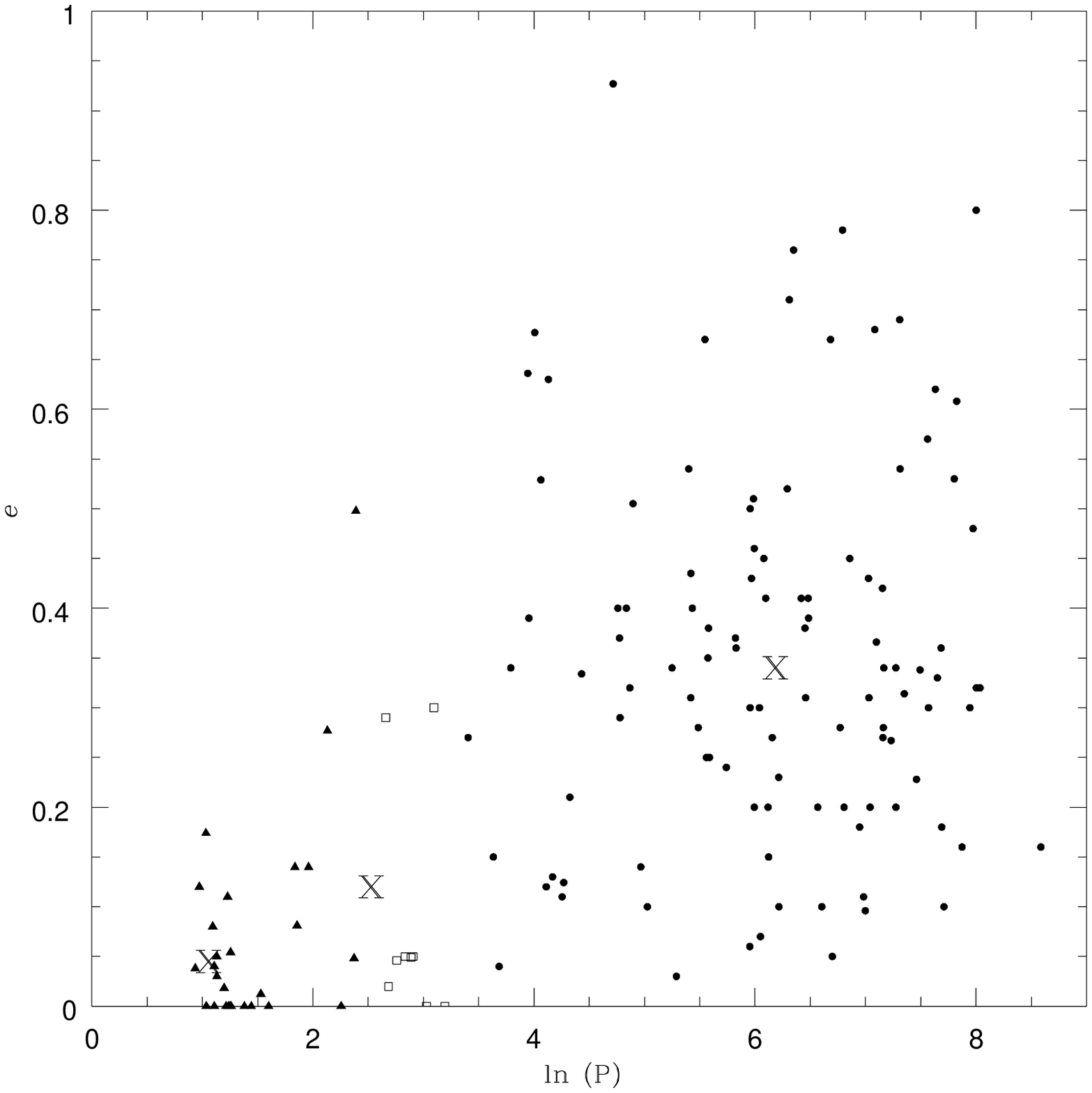}
 \caption{The distribution of exoplanets in the ${\rm ln} P-e$ space.
  There are three crosses
indicating the centers of three clusters. The triangles indicate
the members of the Cluster $Pe_1$, the open squares indicate the
members of the Cluster $Pe_2$, and the full circles indicate the
members of the Cluster $Pe_3$.}
\end{figure}

\clearpage
 \normalsize
 \baselineskip = 24pt

 \noindent
 {\bf Appendix}
 \vskip 0.2truein

 {\small \centerline{ Extra-solar planets data}

    \begin{center}
       \begin{tabular}{|l||l||l|l|l|l|}\hline
 No. & NAME          &  M[.SINI] &  SEM-MAJ.AXIS & PERIOD  & ECC \\
     &               & Jup. mass & (AU)          & days    &     \\\hline\hline
 1   & HD 73256 b    & 1.85      & 0.037         & 2.54863 & 0.038\\\hline
 2   & GJ 436 b      & 0.067     & 0.0278        & 2.6441  & 0.12\\\hline
 3   & 55 Cnc e      & 0.045     & 0.038         & 2.81    & 0.174\\\hline
 4   & 55 Cnc b      & 0.84      &  0.11         & 14.65   & 0.02 \\\hline
 5   & 55 Cnc c      & 0.21      & 0.24          & 44.28   & 0.34\\\hline
 6   & 55 Cnc d      & 4.05      & 5.9           & 5360    & 0.16 \\\hline
 7   & HD 63454 b    & 0.38      &  0.036        & 2.81782 & 0\\\hline
 8   & HD 83443 b    & 0.41      & 0.04          & 2.985   & 0.08\\\hline
 9   & HD 46375 b    & 0.249     & 0.041         & 3.024   & 0.04\\\hline
 10  & TrES-1        & 0.75      & 0.0393        & 3.030065& 0 \\\hline
 11  & HD 179949 b   & 0.84      & 0.045         & 3.093   & 0.05\\\hline
 12  & HD 187123 b   & 0.52      & 0.042         & 3.097   & 0.03\\\hline
 13  & Tau Boo b     & 3.87      & 0.0462        & 3.3128  & 0.018\\\hline
 14  & HD 330075 b   & 0.76      & 0.043         & 3.369   & 0 \\\hline
 15  & HD 88133 b    & 0.29      & 0.046         & 3.415   & 0.11\\\hline
 16  & HD 2638 b     & 0.48      & 0.044         & 3.4442  & 0\\\hline
 17  & BD-10 3166 b  & 0.48      & 0.046         & 3.487   & 0\\\hline
 18  & HD 75289 b    & 0.42      & 0.046         & 3.51    & 0.054\\\hline
 19  & HD 209458 b   & 0.69      & 0.045         &3.52474541& 0\\\hline
 20  & HD 76700 b    & 0.197     & 0.049         & 3.971   & 0 \\\hline
 21  & 51 Peg b      & 0.468     & 0.052         & 4.23077 & 0\\\hline
 22  & Ups And b     & 0.69      & 0.059         & 4.617   & 0.012\\\hline
 23  & Ups And c     & 1.19      & 0.829         & 241.5   & 0.28\\\hline
 24  & Ups And d     & 3.75      & 2.53          & 1284    & 0.27\\\hline
 25  & HD 49674 b    & 0.12      & 0.0568        & 4.948   & 0 \\\hline
 26  & HD 68988 b    & 1.9       & 0.071         & 6.276   & 0.14\\\hline
 27  & HD 168746 b   & 0.23      & 0.065         & 6.403   & 0.081\\\hline
 28  & HD 217107 b   & 1.28      & 0.07          & 7.11    & 0.14 \\\hline
 29  & HD 162020 b   & 13.75     & 0.072         & 8.428198&0.277 \\\hline
 30  & HD 160691 d   & 0.042     & 0.09          & 9.55    & 0 \\\hline
                \end{tabular}
    \end{center}
    }

\newpage
{\small \centerline{ (Continued) Extra-solar planets data}

    \begin{center}
       \begin{tabular}{|l||l||l|l|l|l|}\hline
 No. & NAME          &  M[.SINI] &  SEM-MAJ.AXIS & PERIOD  & ECC \\
     &               & Jup. mass & (AU)          & days    &     \\\hline\hline
 31  & HD 160691 b   & 1.7       & 1.5           & 638     & 0.31\\\hline
 32  & HD 160691 c   & 3.1       & 4.17          & 2986    & 0.8 \\\hline
 33  & HD 130322 b   & 1.08      & 0.088         & 10.724  & 0.048 \\\hline
 34  & HD 108147 b   & 0.41      & 0.104         & 10.901  & 0.498 \\\hline
 35  & HD 38529 b    & 0.78      & 0.129         & 14.309  & 0.29 \\\hline
 36  & HD 38529 c    & 12.7      & 3.68          & 2174.3  & 0.36 \\\hline
 37  & Gl 86 b       & 4         & 0.11          & 15.78   & 0.046 \\\hline
 38  & HD 99492 b    & 0.112     & 0.119         & 17.038  & 0.05 \\\hline
 39  & HD 27894 b    & 0.62      & 0.122         & 17.991  & 0.049 \\\hline
 40  & HD 195019 b   & 3.43      & 0.14          & 18.3    & 0.05 \\\hline
 41  & HD 6434 b     & 0.48      & 0.15          & 22.09   & 0.3 \\\hline
 42  & HD 192263 b   & 0.72      & 0.15          & 24.348  & 0  \\\hline
 43  & Gliese 876 c  & 0.56      & 0.13          & 30.1    & 0.27 \\\hline
 44  & Gliese b      & 1.98      & 0.21          & 61.02   & 0.12 \\\hline
 45  & HD 102117 b   & 0.14      & 0.149         & 20.67   & 0  \\\hline
 46  & HD 11964 c    & 0.11      & 0.23          & 37.82   & 0.15 \\\hline
 47  & HD 11964 b    & 0.7       & 3.17          & 1940    & 0.3 \\\hline
 48  & rho CrB b     & 1.04      & 0.22          & 39.845  & 0.04 \\\hline
 49  & HD 74156 b    & 1.86      & 0.294         & 51.643  & 0.636 \\\hline
 50  & HD 117618 b   & 0.19      & 0.28          & 52.2    & 0.39 \\\hline
 51  & HD 37605 b    & 2.4       & 0.26          & 55.02   & 0.677 \\\hline
 52  & HD 168443 b   & 7.7       & 0.29          & 58.116  & 0.529 \\\hline
 53  & HD 168443 c   & 16.9      & 2.85          & 1739.5  & 0.228 \\\hline
 54  & HD 3651 b     & 0.2       & 0.284         & 62.23   & 0.63 \\\hline
 55  & HD 121504 b   & 0.89      & 0.32          & 64.6    & 0.13  \\\hline
 56  & HD 101930 b   & 0.3       & 0.302         & 70.46   & 0.11 \\\hline
 57  & HD 178911 B b & 6.292     & 0.32          & 71.487  & 0.1243 \\\hline
 58  & HD 16141 b    & 0.23      & 0.35          & 75.56   & 0.21  \\\hline
 59  & HD 114762 b   & 11        & 0.3           & 84.03   & 0.334 \\\hline
 60  & HD 80606 b    & 3.41      & 0.439         & 111.78  & 0.927\\\hline
             \end{tabular}
    \end{center}
    }

\newpage
{\small \centerline{ (Continued) Extra-solar planets data}

    \begin{center}
       \begin{tabular}{|l||l||l|l|l|l|}\hline
 No. & NAME          &  M[.SINI] &  SEM-MAJ.AXIS & PERIOD  & ECC \\
     &               & Jup. mass & (AU)          & days    &     \\\hline\hline
 61  & 70 Vir b      & 7.44      & 0.48          & 116.689 & 0.4  \\\hline
 62  & HD 216770 b   & 0.65      & 0.46          & 118.45  & 0.37 \\\hline
 63  & HD 52265 b    & 1.13      & 0.49          & 118.96  & 0.29 \\\hline
 64  & HD 34445 b    & 0.58      & 0.51          & 126.    & 0.4  \\\hline
 65  & HD 208487 b   & 0.45      & 0.49          & 130     & 0.32 \\\hline
 66  & HD 93083 b    & 0.37      & 0.477         & 143.58  & 0.14 \\\hline
 67  & GJ 3021 b     & 3.21      & 0.49          & 133.82  & 0.505 \\\hline
 68  & HD 37124 b    & 0.75      & 0.54          & 152.4   & 0.1 \\\hline
 69  & HD 37124 c    & 1.2       & 2.5           & 1495    & 0.69 \\\hline
 70  & HD 73526 b    & 3         & 0.66          & 190.5   & 0.34 \\\hline
 71  & HD 104985 b   & 6.3       & 0.78          & 198.2   & 0.03 \\\hline
 72  & HD 82943 b    & 0.88      & 0.73          & 221.6   & 0.54 \\\hline
 73  & HD 82943 c    & 1.63      & 1.16          & 444.6   & 0.41  \\\hline
 74  & HD 169830 b   & 2.88      & 0.81          & 225.62  & 0.31  \\\hline
 75  & HD 169830 c   & 4.04      & 3.6           & 2102    & 0.33 \\\hline
 76  & HD 8574 b     & 2.23      & 0.76          & 228.8   & 0.4\\\hline
 77  & HD 202206 b   & 17.4      & 0.883         & 225.87  & 0.435 \\\hline
 78  & HD 202206 c   & 2.44      & 2.55          & 1383.4  & 0.267\\\hline
 79  & HD 89744 b    & 7.99      & 0.89          & 256.6   & 0.67 \\\hline
 80  & HD 134987 b   & 1.58      & 0.78          & 260     & 0.25 \\\hline
 81  & HD 40979 b    & 3.32      & 0.811         & 267.2   & 0.25 \\\hline
 82  & HD 12661 b    & 2.3       & 0.83          & 263.6   & 0.35 \\\hline
 83  & HD 12661 c    & 1.57      & 2.56          & 1444.5  & 0.2 \\\hline
 84  & HD 150706 b   & 1         & 0.82          & 264.9   & 0.38 \\\hline
 85  & HR 810 b      & 1.94      & 0.91          & 311.288 & 0.24 \\\hline
 86  & HD 142 b      & 1.36      & 0.98          & 338     & 0.37 \\\hline
 87  & HD 92788 b    & 3.8       & 0.94          & 340     & 0.36\\\hline
 88  & HD 28185 b    & 5.6       & 1             & 385     & 0.06 \\\hline
 89  & HD 196885 b   & 1.84      & 1.12          & 386     & 0.3 \\\hline
 90  & HD 142415 b   & 1.62      & 1.05          & 386.3   & 0.5 \\\hline
           \end{tabular}
    \end{center}
    }

\newpage
{\small \centerline{ (Continued) Extra-solar planets data}

    \begin{center}
       \begin{tabular}{|l||l||l|l|l|l|}\hline
 No. & NAME          &  M[.SINI] &  SEM-MAJ.AXIS & PERIOD  & ECC \\
     &               & Jup. mass & (AU)          & days    &     \\\hline\hline
  91 & HD 177830 b   & 1.28      & 1             & 391     & 0.43 \\\hline
  92 & HD 154857 b   & 1.8       & 1.11          & 398     & 0.51 \\\hline
  93 & HD 108874 b   & 1.65      & 1.07          & 401     & 0.2 \\\hline
  94 & HD 4203 b     & 1.65      & 1.09          & 400.944 & 0.46 \\\hline
  95 & HD 128311 b   & 2.58      & 1.02          & 420     & 0.3 \\\hline
  96 & HD 27442 b    & 1.28      & 1.18          & 423.841 & 0.07 \\\hline
  97 & HD 210277 b   & 1.28      & 1.097         & 437     & 0.45 \\\hline
  98 & HD 19994 b    & 2         & 1.3           & 454     & 0.2  \\\hline
 99  & HD 188015 b   & 1.26      & 1.19          & 456.46  & 0.15\\\hline
 100 & HD 13189 b    & 14        & 1.85         & 471.6   & 0.27\\\hline
 101  & HD 20367 b    & 1.07      & 1.25          & 500     & 0.23\\\hline
 102  & HD 114783 b   & 0.9       & 1.2           & 501     & 0.1\\\hline
 103  & HD 147513 b   & 1         & 1.26          & 540.4   & 0.52\\\hline
 104  & HIP 75458 b   & 8.64      & 1.34          & 550.651 & 0.71\\\hline
 105  & HD 222582 b   & 5.11      & 1.35          & 572     & 0.76\\\hline
 106  & HD 65216 b    & 1.21      & 1.37          & 613.1   & 0.41\\\hline
 107  & HD 183263 b   & 3.69      & 1.52          & 634.23  & 0.38\\\hline
 108  & HD 141937 b   & 9.7       & 1.52          & 653.22  & 0.41\\\hline
 109  & HD 41004A b   & 2.3       & 1.31          & 655     & 0.39\\\hline
 110  & HD 47536 b    & 7.32      & 1.93          & 712.13  & 0.2\\\hline
 111  & HD 23079 b    & 2.61      & 1.65          & 738.459 & 0.1\\\hline
 112  & 16 CygB b     & 1.69      & 1.67          & 798.938 & 0.67\\\hline
 113  & HD 4208 b     & 0.8       & 1.67          & 812.197 & 0.05\\\hline
 114  & HD 114386 b   & 0.99      & 1.62          & 872     & 0.28\\\hline
 115  & HD 45350 b    & 0.98      & 1.77          & 890.76  & 0.78\\\hline
 116  & gamma Cephei b& 1.59      & 2.03          & 902.96  & 0.2\\\hline
 117  & HD 213240 b   & 4.5       & 2.03          & 951     & 0.45\\\hline
 118  & HD 10647 b    & 0.91      & 2.1           & 1040    & 0.18\\\hline
 119  & HD 10697 b    & 6.12      & 2.13          & 1077.906& 0.11\\\hline
 120  & 47 Uma b      & 2.41      & 2.1           & 1095    & 0.096\\\hline
          \end{tabular}
    \end{center}
    }

\newpage
{\small \centerline{ (Continued) Extra-solar planets data}

    \begin{center}
       \begin{tabular}{|l||l|l|l|l|l|}\hline
 No.  & NAME          &  M[.SINI] &  SEM-MAJ.AXIS & PERIOD  & ECC \\
      &               & Jup. mass & (AU)          & days    &     \\\hline\hline
 121  & HD 190228 b   & 4.99      & 2.31          & 1127    & 0.43\\\hline
 122  & HD 114729 b   & 0.82      & 2.08          & 1131.478& 0.31\\\hline
 123  & HD 111232 b   & 6.8       & 1.97          & 1143    & 0.2\\\hline
 124  & HD 2039 b     & 4.85      & 2.19          & 1192.582& 0.68\\\hline
 125  & HD 136118 b   & 11.9      & 2.335         & 1209.6  & 0.366\\\hline
 126  & HD 50554 b    & 4.9       & 2.38          & 1279    & 0.42\\\hline
 127  & HD 196050 b   & 3         & 2.5           & 1289    & 0.28\\\hline
 128  & HD 216437 b   & 2.1       & 2.7           & 1294    & 0.34\\\hline
 129  & HD 216435 b   & 1.49      & 2.7           & 1442.919& 0.34\\\hline
 130  & HD 106252 b   & 6.81      & 2.61          & 1500    & 0.54\\\hline
 131  & HD 23596 b    & 7.19      & 2.72          & 1558    & 0.314\\\hline
 132  & 14 Her b      & 4.74      & 2.8           & 1796.4  & 0.338\\\hline
 133  & HD 142022     & 4.4       & 2.8           & 1923    & 0.57\\\hline
 134  & HD 39091 b    & 10.35     & 3.29          & 2063.818& 0.62\\\hline
 135  & HD 72659 b    & 2.55      & 3.24          & 2185    & 0.18\\\hline
 136  & HD 70642 b    & 2         & 3.3           & 2231    & 0.1\\\hline
 137  & HD 33636 b    & 9.28      & 3.56          & 2447.292& 0.53\\\hline
 138  &Epsilon Eridanib& 0.86     & 3.3           & 2502.1  & 0.608\\\hline
 139  & HD 117207 b   & 2.06      & 3.78          & 2627.08 & 0.16\\\hline
 140  & HD 30177 b    & 9.17      & 3.86          & 2819.654& 0.3\\\hline
 141  & HD 50499 b    & 1.84      & 4.403         & 2990    & 0.32\\\hline
 142  & Gl 777A b     & 1.33      & 4.8           & 2902    & 0.48\\\hline
 143  & HD 89307 b    & 2.73      & 4.15          & 3090    & 0.32\\\hline
    \end{tabular}
    \end{center}
    }


\clearpage

\begin{thebibliography}{21}

\bibitem{A} Armitage, P. J., Livio, M., Lubow, S. H., Pringle, J. E.,
2002, MNRAS, 334, 248

\bibitem{B1} Baggaley, W. J., Galligan, D. P., 1997,
    Planetary and Space Science, 45, 865

\bibitem{G1} Galligan, D. P., 2003a, MNRAS, 340, 893

\bibitem{G2} Galligan, D. P., 2003b, MNRAS, 340, 899

\bibitem{Go} Gower, J. C., Ross, G. J. S., 1969, Appl. Stat., 18, 54

\bibitem{h1} Hartigan, J.A., 1967, J. Am. Stat. Ass., 62, 1140

\bibitem{h2} Hartigan, J.A., 1975, {\it Clustering Algorithms}
(Wiley, New York)

\bibitem{Ji1} Ji, J., Kinoshita, H., Liu, L., Li, G. 2003,
ApJ, 585, L139

\bibitem{Ji2} Ji, J., Li, G., Liu, L. 2002, ApJ, 572, 1041

\bibitem{Jiang}  Jiang, I.-G., Ip, W.-H., 2001, A\&A,  367, 943

\bibitem{JiangY}  Jiang, I.-G., Ip, W.-H., Yeh, L.-C., 2003, ApJ, 582, 449

\bibitem{JiangY1}  Jiang, I.-G., Yeh, L.-C. 2004a, AJ, 128, 923

\bibitem{JiangY2}  Jiang, I.-G., Yeh, L.-C. 2004b,
Int. J. Bifurcation and Chaos, 14, 3153

\bibitem{Jw} Johnson, R. A., Wichern, D. W., 1988, {\it Applied Multivariate
 Statistical Analysis} (Prentice-Hall, Inc.)

\bibitem{mac} MacQueen, J., 1967, {\it Proceeding of the Fifth Berkeley 
Symposium on Mathematical Statistics and Probability} 
(University of California Press, Berkeley), 281


\bibitem{P} P\"atzold, M., Rauer, H., 2002, ApJ, 568, L117

\bibitem{T1} Tabachnik, S., Tremaine, S., 2002, MNRAS, 335, 151

\bibitem{T2} Trilling, D. E., Lunine, J. I., Benz, W., 2002,
A\&A, 394, 241

\bibitem{V1} Veras, D., Armitage, P. J., 2004, MNRAS, 347, 613

\bibitem{Ya} Yang, M.-S., Wu, K.-L., 2004, IEEE Trans. PAMI, 26, 434

\bibitem{Y1} Yeh, L.-C., Jiang, I.-G., 2001,
ApJ, 561, 364

\bibitem{Z0} Zakamska, N. L., Tremaine, S., 2004, AJ, 128, 869

\bibitem{Z1} Zappala, V., Bendjoya, P., Cellino, A., Farinella, P.,
Froeschle, C., 1995, Icarus, 116, 291

\bibitem{Z2} Zucker, S., Mazeh, T., 2002, ApJ, 568, L113


\end{thebibliography}
\end{document}